%% file: potts090603.tex
\documentclass{JHEP3}
\usepackage{amsmath}
\usepackage{graphicx}
\usepackage{amsfonts}
\usepackage{amssymb}
\usepackage[all]{xy}\CompileMatrices

\newcommand{\bea}{\begin{eqnarray}}
\newcommand{\eea}{\end{eqnarray}}
\newcommand{\beq}{\begin{equation}}
\newcommand{\eeq}{\end{equation}}
\newcommand{\nn}{\nonumber}

\newcommand{\lishi}{\langle\!\langle}
\newcommand{\rishi}{\rangle\!\rangle}

\newcommand{\abar}{{\overline a}}
\newcommand{\alphabar}{{\overline \alpha}}
\newcommand{\Lbar}{{\overline L}}
\newcommand{\Fbar}{{\hat F}}
\newcommand{\Ibar}{{\hat I}}
\newcommand{\Ebar}{{\hat \epsilon}}

\newcommand{\C}[1]{{\mathcal{#1}}}
\newcommand{\B}[1]{{\mathbf{#1}}}
\newcommand{\sst}{\scriptscriptstyle {}}

\newcommand{\allint}{{\mathbb Z}}
\newcommand{\half}{{\frac{1}{2}}}
\newcommand{\threehalves}{{\frac{3}{2}}}
\newcommand{\threequarters}{{\frac{3}{4}}}
\newcommand{\cbytwelve}{{\frac{c}{12}}}
\newcommand{\Ham}{\half(L_0+\Lbar_0-\cbytwelve)}
\newcommand{\ket}[1]{\vert #1\rangle}

\newcommand{\ishiket}[1]{\vert\, #1\,\rishi}
\newcommand{\ishibra}[1]{\lishi\, #1\,\vert}

\preprint{OUTP-03/12P\\ HIP-2003-31/TH}

\title{Free boson formulation of boundary states in $W_3$ minimal models and the critical Potts model}
\author{Alexandre F. Caldeira\\Department of Physics, University of Oxford \\
Theoretical Physics,\\
1 Keble Road,\\
 Oxford OX1 3NP, UK\\
E-mail: \email{caldeira@thphys.ox.ac.uk}}
\author{Shinsuke Kawai\\Helsinki Institute of Physics,\\
P.O.Box 64,\\
FIN-00014 University of Helsinki,\\
Finland\\
E-mail: \email{shinsuke.kawai@helsinki.fi}}
\author{John F. Wheater\\ Department of Physics, University of Oxford \\
Theoretical Physics,\\
1 Keble Road,\\
 Oxford OX1 3NP, UK\\
E-mail: \email{j.wheater@physics.ox.ac.uk}}
\abstract{
We  develop a Coulomb gas formalism for boundary conformal field theory having
a $W$ symmetry and illustrate its operation using
 the three state Potts model. We find that there are 
free-field representations for six $W$ conserving boundary states,
which yield the  fixed and mixed physical boundary
conditions, and two $W$ violating boundary states which yield the 
free and new boundary conditions. Other $W$ violating boundary states can be constructed 
but they decouple from the rest of the theory. Thus we have a complete free-field realization 
of the known boundary states of the three state Potts model.
We then use the formalism to calculate boundary correlation functions in various cases. 
We find that the conformal blocks arising when the
two point function of $\phi_{2,3}$ is calculated in the presence of free and new boundary
 conditions are indeed the last two solutions of the sixth order differential equation generated by the singular vector.
}
\dedicated{Dedicated to the memory of Ian I. Kogan, 1958-2003}

\begin{document}

\section{Introduction}

The  Coulomb gas formalism \cite{Dotsenko:1984nm, Dotsenko:1985ad} provides a powerful 
method for calculating correlation functions and conformal blocks in minimal rational  
conformal field theories (CFTs) without having to solve complicated differential equations.
 Boundary CFTs have been of great interest since Cardy's famous paper \cite{Cardy:1984bb}
and  recently it has been shown that free-field  representations may be extended from  bulk CFTs to 
systems with boundary(ies) \cite{Kawai:2002vd,Kawai:2002pz} in the case of the Virasoro diagonal 
minimal models. The boundary states appear as coherent states in the free-field formalism.
As in the bulk case these methods again provide a mechanism for calculating 
correlation functions, this time in the presence of a boundary.

The three state Potts model is of particular interest because the conformal field theory describing
its critical point is the simplest in which there is a higher dimensional chiral operator of integer
dimension. The resulting extended symmetry, generated by  a $W_3$ algebra, has important consequences.
In particular the theory is diagonal in the bigger $W_3$ algebra but not in the smaller Virasoro algebra.
Such theories can be represented by a multi-component Coulomb gas formalism    \cite{Fateev:1987vh}.
A great deal is known about the boundary states in this model. There are six states originally found 
by Cardy in which the $W$ current is conserved  at the boundary   \cite{Cardy:1989ir} and which 
correspond to fixed and mixed boundary conditions in the spin system. In addition there are
known to be  two more
states in which the $W$ current is not conserved. One of them is the state corresponding to free boundary conditions
on the spin system, which was of course expected to exist, and the other is the so-called
new boundary condition; these were found in  \cite{Affleck:1998nq}. Subsequently it was shown in
 \cite{Fuchs:1998qn} that the set of eight states is in some sense complete and in
 \cite{Behrend:1998mu} it was suggested that this completeness amounts to the set of boundary
conditions for which the model is an integrable system.  Clearly none of these boundary states
can correspond to boundary conditions on the spin system
 which are not conformally invariant.

The purpose of this paper is to show how the multi-component Coulomb gas formalism 
may be used to provide
a free-field description of the boundary states in theories with a $W$ algebra using
  the three state  Potts model to illustrate the method. 
We start in section 2 by briefly reviewing those aspects of the standard formalism 
that we need.
In section 3 we record in detail the representations of the primary fields for the Potts model 
and in section 4 explain how to construct the free-field representations for Cardy's boundary states. Section 5 contains some example calculations of correlation functions in the
presence of boundaries; these act as a useful cross-check that the states we have constructed are correct.

Finally in section 6 we discuss some open issues and some useful results are recorded in the appendices.

\section{Free-field construction of CFTs with $W$ algebra}

The usual Coulomb gas formalism \cite{Dotsenko:1984nm, Dotsenko:1985ad} can be extended to CFTs with a
larger symmetry than the Virasoro algebra by introducing a multiple component
scalar field \cite{Fateev:1987vh} 
 $\Phi^j(z,\bar z)$, $j=1\ldots r$, which is a 
vector in the root space of a finite dimensional Lie Algebra $\mathcal A$ of rank
$r$. In practice we will be considering the simplest case in this paper so
we will confine ourselves to the algebra $A_r$ from the outset.
 Let us first fix some notation. The simple roots will be denoted
$e_j, j=1\ldots r$, and the corresponding dual weights $\omega_j, j=1\ldots r$.
We will use ``$\cdot$'' to denote multiplication of vectors and matrices in the 
root space. So the scalar product of two vectors
$u$ and $v$ in the root space will be written $u\cdot v$, the product of two matrices
$m_1\cdot m_2$ and so on. The simple roots and
dual weights then satisfy
\begin{equation}
e_j \cdot e_j=2,\quad  e_j \cdot e_{j+1}=-1, \quad e_j \cdot\omega_i =\delta_{i,j}.
\end{equation}
The Weyl vector $\rho$ is defined as
\begin{equation}
\rho=\sum_{j=1}^r  e_j = \sum_{j=1}^r \omega_j 
\end{equation}
and the fundamental weights $h_K, K=1\ldots r+1,$ satisfy
\begin{eqnarray}
h_1&=&\omega_1,\nonumber\\
h_K-h_{K+1}&=&e_K.
\end{eqnarray}
We denote the  Weyl group of  $\mathcal A$ by  $\mathcal W$, an element of
it by  $w$, and let $\varepsilon_w=\det w$; it is convenient also to define the shifted Weyl transformation by
\begin{equation}
 w^{\ast}(\gamma)=w(\gamma+\rho)-\rho.
\end{equation}
In the particular case of $A_2$ we have the useful relations
\begin{eqnarray}
e_1&=&2\omega_1-\omega_2,\nonumber\\
e_2&=&2\omega_2-\omega_1,\nonumber\\
h_2&=&\omega_2-\omega_1,\nonumber\\
h_3&=&-\omega_2.
\label{rootrels}\end{eqnarray}
The Weyl group is of order 6 and is generated by $w_1$ (reflections in
$e_1$) and $w_2$ (reflections in
$e_2$) 
\begin{equation}
\mathcal W=\{w_0=I,w_1,w_2,w_3=w_1w_2w_1\equiv w_\rho,w_4= w_2w_1,w_5=w_1w_2\}.
\end{equation}

The action for $\Phi$ takes the usual form
\begin{equation}
\mathcal{S}[\Phi]=\frac{1}{8\pi}\int d^{2}z\sqrt{g}\,\left(\partial_{\mu}%
\Phi\cdot \partial^{\mu}\Phi+4 i\alpha_{0}\rho\cdot\Phi R\right), \label{s}%
\end{equation}
where $R$ is the scalar curvature, $g$ the metric, and $\alpha_0$ a constant.
We now split $\Phi$ into a holomorphic component $\phi(z)$ and an 
anti-holomorphic component $\bar\phi(\bar z)$.
The field $\phi$ has mode expansion
\begin{equation}
\phi^{j}(z)=\phi_{0}^{j}-ia_{0}^{j}\ln z+i\sum_{n\neq0}\frac{a_{n}^{j}%
}{n}z^{-n}, \label{eq. phimode}%
\end{equation}
and similarly for  $\bar\phi$.
Canonical quantization gives the usual commutation relations
\begin{align}
\lbrack a_{m}^{j},a_{n}^{l}]  &  =m\delta^{jl}\delta_{m+n,0}%
,\label{eqn:heisenberg}\\
\lbrack\phi_{0}^{j},a_{0}^{l}]  &  =i\delta^{jl}. \label{heisemberg2}%
\end{align}
Variation of the action with respect to the metric yields the
energy-momentum tensor
\begin{equation}
T(z)=-2\pi T_{zz}=-\frac{1}{2}:\partial\phi\cdot\partial\phi
:+2\,i\,\alpha_{0}\,\rho\cdot\partial^{2}\phi, \label{T}%
\end{equation}
which has the usual expansion 
\begin{equation}T(z)=\sum_{n\in{\mathbb Z}}%
L_{n}z^{-n-2},\end{equation}
where the operators
\begin{align}
L_{n}  &  =\frac{1}{2}\sum_{m\in{\mathbb Z}}  :a_{m}\cdot a_{n-m}:
  -2\alpha_{0}(n+1)\rho\cdot a_{n}
\end{align}
obey the Virasoro algebra with
 central charge
\begin{equation}
c=r-48\ \alpha_{0}^{2}\rho^2.
\end{equation}

Fock spaces $\C F_\alpha$ are  labeled by a vacuum $|\alpha\rangle$, 
which is an
eigenvector of the $a_{0}^{j}$ operator, and annihilated by the positive
modes
\begin{eqnarray}a_{0}^{j}|\alpha\rangle&=&\alpha^{j}|\alpha\rangle,\nonumber\\
a_{n}^{j}|\alpha\rangle&=&0,\quad n>0.
\end{eqnarray}
The Fock space is formed by applying the creation
operators to the vacuum,
\begin{equation}
 a_{-n_{1}}^{j_{1}}a_{-n_{2}}^{j_{2}}...a_{-n_{p}%
}^{j_{p}}|\alpha\rangle,
\end{equation}
 and different Fock spaces are related by 
\begin{equation}
 e^{i\beta\cdot\phi_{0}}|\alpha\rangle
=|\beta+\alpha\rangle,
\end{equation}
 as can be  checked using the commutation
relations (\ref{heisemberg2}).
The  physical Hilbert space
  of the  theory is much smaller than the direct sum of the Fock spaces
and is given by the cohomology of a nilpotent BRST operator \cite{Felder:1989zp}.

The chiral vertex operators for this theory are defined by 
generalizing the vertex operators of the $U(1)$ Coulomb gas formalism,
\begin{equation}
V_{\alpha}(z)=:e^{i\alpha\cdot\phi(z)}:.
\end{equation}
It is straightforward to check, by computing the OPE $T(z)V_{\alpha}(z')$, that vertex
operators are primary fields with conformal dimensions given by 
\begin{equation}
h(\alpha
)=\frac{1}{2}\alpha\cdot(\alpha-4\alpha_{0}\rho).\label{Halpha}
\end{equation}
Note that
\begin{equation}
h(\alpha
)=h(4\alpha_{0}\rho- \alpha  ).\label{Hdouble}
\end{equation}

This generalized free-field construction contains an extra symmetry 
which  corresponds to the presence of the
higher dimensional conserved chiral primary fields found in theories with
a $W$ symmetry (see \cite{Bilal:1991eu, Bouwknegt:1993wg} for reviews).
 The new chiral fields can
be generated systematically in the following way.
 First define the generating 
functional $\mathcal D_N$, where $N=r+1$, by
\beq
(2i\alpha_{0})^{N}\mathcal{D}_{N}    =\,:\prod_{K=1}^{N}\left(  2i\alpha
_{0}\partial_{z}+h_{K}\cdot\partial\phi(z)\right):\,.
\eeq
This can be evaluated to get 
\begin{equation}
\mathcal{D}_{N}=\partial^{N}+\sum_{K=1}^{N}(2i\alpha_{0})^{-K}u_{K}[\phi(z)]\;
\partial^{N-K},
\end{equation}
where it turns out that $u_1=0$ and $u_2=T(z)$ and the higher $W$ currents are given by
\begin{eqnarray}
W_K=u_{K}+b\partial u_{K-1}&+&b'\partial^2 u_{K-2}+\ldots\nonumber \\
&+&b'' u_2u_{K-2}+\ldots\nonumber \\
&\vdots&\qquad\qquad\qquad,
\end{eqnarray}
where $b$ etc are constants. For our purposes in this paper it is sufficient to
know the generating functional  $\mathcal{D}_{N}$ and 
the zero-mode of $W_3(z)$ for $A_2$ which is, 
\bea W_0=ix^{-1}\Big(&& \sum_{l+m+n=0}h_1\cdot a_l\,h_2\cdot a_m\,h_3\cdot a_n\nn\\
&&+2\alpha_0\sum_m [(1-m) h_3\cdot a_m\,h_1\cdot a_{-m} +2h_2\cdot a_m\,h_1\cdot a_{-m}]\nn\\
&&+8\alpha_0^2\,h_1\cdot a_0 +2\alpha_0 L_0\Big),\label{Wzero}
\eea
where $x=\sqrt{(22+5c)/48}$.

Screening operators $Q_{\alpha}$ are defined  by analogy with the 
one component Coulomb gas formalism as
\begin{equation}
Q_{\alpha}=\oint dz:e^{i\alpha\cdot\phi(z)}:, \label{Q}%
\end{equation}
where now  $\alpha$  is any vector living on the $r-1$ dimensional manifold
 $h(\alpha)=1$.
Since for any such $\alpha$
\begin{equation} T(z) V_\alpha(w)=\partial_w(\ldots)+{\rm regular~terms},
\end{equation}
the $Q_{\alpha}$
 commute automatically with the Virasoro algebra,
\begin{equation}
\lbrack L_{n},Q_{\alpha}]=0.
\end{equation}
However  in general  the $Q_{\alpha}$ do not commute with the $W$ algebra and it can be shown (see appendix A) that
the only ones which do are (in the particular case of $A_2$)
\begin{eqnarray}
Q^{(1)}_\pm&=&\oint dz:e^{i\alpha_\pm e_1\cdot\phi(z)}:,\nonumber\\
Q^{(2)}_\pm&=&\oint dz:e^{i\alpha_\pm e_2\cdot\phi(z)}:.
\label{screeners}\end{eqnarray}

\section{Free-field approach to the three state Potts model}
 The three state Potts model  is a minimal
 model that is not
diagonal in the ``small'' Virasoro algebra but becomes diagonal in the larger
$W_{3}$ algebra. To find its representation in terms of the $A_2$ Coulomb gas
 we have to find
irreducible highest weight representations for which the Verma modules are
completely degenerate, that is to say  have as many singular vectors as possible.
This can be done by considering a theory,  $W_{3}(p,p^{\prime})$, parametrized by
relative primes $p$ and $p\prime$  where
\begin{eqnarray} 2\alpha_{0}&=&\frac{p^{\prime}-p}{\sqrt{pp\prime}},\nonumber\\
\alpha_{+}&=&\frac{p\prime}{\sqrt{pp\prime}},\nonumber\\
\alpha_{-}&=&-\frac{p}{\sqrt{pp\prime}},
\end{eqnarray}
for which the central charge is 
\begin{equation}
c_{(p,p^{\prime})}^{(3)}=2-24\frac{\left(  p^{\prime}-p\right)
^{2}}{pp\prime}.
\end{equation}
Choosing  $p=4$, $p^{\prime}=5$ gives the correct central charge, $4/5$, for the Potts
model. The field content is represented
 by highest weight states through
\begin{equation}
\alpha=-\alpha_{+}\lambda_{+}-\alpha_{-}\lambda_{-},%
\end{equation}
where $\lambda_\pm$ are dominant weights of the algebra
parametrized by two pairs of
integers, $(r_{1},r_{2})\,$ and $(s_{1},s_{2})$,
\begin{equation}
\lambda_{+}=r_{1}\omega_{1}+r_{2}\omega_{2},\qquad\lambda_{-}=s_{1}\omega
_{1}+s_{2}\omega_{2},
\end{equation}
and $0\le s_{1,2}\le 3$ and $0\le r_{1,2} \le 2$.
For each primary field $F$ there are six $\alpha\,$s which all lie in the same
Weyl orbit and can be written
\begin{eqnarray}
\alpha_{i}(F)&=&-2\alpha_0w_i^*\left(-p\Lambda_-(F)\right),\quad
i=0,\ldots 5,\nonumber\\
\Lambda_-(F)&=&s_{1}\omega_{1}+s_{2}\omega_{2},\nn\\
\Lambda_+(F)&=&r_{1}\omega_{1}+r_{2}\omega_{2},\label{primarydefn}
\end{eqnarray}
where $r_{1,2}$ and $s_{1,2}$ are read off from the first column of Table 1. An equivalent
way of generating these is to compute the $\alpha\,$s from all three columns
of Table 1 and form the sets
\begin{eqnarray}
I,\,\epsilon &:&\{\alpha,4\alpha_{0}\rho-\alpha\},\nonumber\\
\sigma,\,\psi&:& \{\alpha_{\sigma,\psi},4\alpha_{0}\rho-\alpha_{\sigma^\dag,\psi^\dag}\},\nonumber\\
\sigma^\dag,\,\psi^\dag&:& \{\alpha_{\sigma^\dag,\psi^\dag},4\alpha_{0}\rho-\alpha_{\sigma,\psi}\}.
\end{eqnarray}
The conformal dimensions of the fields obtained this way using (\ref{Halpha})
are the same as for the fields in the three state Potts model which are shown in Table 2.
This six-fold redundancy in the choice of $\alpha$s for the primary fields  in the bulk theory
is a consequence of the property (\ref{Hdouble}) and the $Z_3$ symmetry of the root diagram
for $A_2$ which is in turn responsible for the existence of the chiral $W_3$ algebra. This changes when we 
consider boundary conditions which only preserve the Virasoro symmetry and not the full $W$ symmetry. 
As we will see in section 4, only the Felder complexes (discussed below) built on the first 
column of Table 1 yield $W$ violating boundary states which do not decouple from the bulk theory.
\DOUBLETABLE{
\begin{tabular}
[c]{c|c|c|c}\hline
& \multicolumn{3}{c}{$(r_{1},r_{2})$}\\\cline{2-4}%
$(s_{1},s_{2})$ & $(0,0)$ & $(0,1)$ & $(1,0)$\\\hline\hline
$(0,0)$ & $I$ & $\psi^{\dag}$ & $\psi$\\\hline
$(0,1)$ & $\sigma^{\dag}$ & $\sigma$ & $\epsilon$\\\hline
$(1,0)$ & $\sigma$ & $\epsilon$ & $\sigma^{\dag}$\\\hline
$(0,2)$ & $\psi$ & $I$ & $\psi^{\dag}$\\\hline
$(1,1)$ & $\epsilon$ & $\sigma^{\dag}$ & $\sigma$\\\hline
$(2,0)$ & $\psi^{\dag}$ & $\psi$ & $I$\\\hline
\end{tabular}}
{\begin{tabular}
[c]{c|c|c|c|c|c}\hline
$I$ & $\sigma$ & $\sigma^{\dag}$&$\psi$& $\psi^{\dag}$ &  $\epsilon$\\\hline
$0$ & $\frac{1}{15}$& $\frac{1}{15}$& $\frac{2}{3}$& $\frac{2}{3}$& $\frac{2}{5}$\\\hline
\end{tabular}}
{ The representations for the primary fields in the 3-state Potts model.}
{ The conformal dimensions of the primary fields in the 3-state Potts model.}

The physical Hilbert space is determined by the BRST structure first elucidated by Felder
\cite{Felder:1989zp} and extended to the $W_3$ case in \cite{Bouwknegt:1990xa,Mizoguchi:1992vt}.
 In this construction it is more convenient to write $\alpha(F)$ in the form
\beq \alpha(F)=\alpha_-((1-m_1)\omega_1+(1-m_2)\omega_2)+\alpha_+((1-n_1)\omega_1+(1-n_2)\omega_2)
\eeq
and to denote the corresponding Fock space by $\C F_\alpha=\C F^{\B m}_{n_1,n_2}$ where $\B m=(m_1,m_2)$.
The first two BRST charges are
\bea Q^{(1)}_k&=&e^{i(k-1)\theta}\,\frac{\sin k\theta}{\sin\theta}
\prod_{j=1..k}\oint dz_j V_{\alpha_+ e_1}(z_j),\nn\\
 Q^{(2)}_k&=&e^{i(k-1)\theta}\,\frac{\sin k\theta}{\sin\theta}\prod_{j=1..k}\oint dz_j V_{\alpha_+ e_2}(z_j),\eea
with the usual Felder contour prescription and where $\theta=2\pi\alpha_+^{\,2}$.
These do not (anti-) commute and it is necessary \cite{Bouwknegt:1990xa} to introduce a third charge
$Q^{(3)}_k$ having the property
\bea
Q^{(3)}_{k_1}Q^{(1)}_{k_2}&=&Q^{(1)}_{k_1+k_2}Q^{(2)}_{k_1}.
\eea
All the $Q^{(i)}$ commute with the Virasoro algebra by construction
and  their action  on the Fock spaces is
\bea Q^{(1)}_k\C F^{\B m}_{n_1,n_2}&=&\C F^{\B m}_{n_1-2k,n_2+k}\,,\nn\\
Q^{(2)}_k\C F^{\B m}_{n_1,n_2}&=&\C F^{\B m}_{n_1+k,n_2-2k}\,,\nn\\
Q^{(3)}_k\C F^{\B m}_{n_1,n_2}&=&\C F^{\B m}_{n_1-k,n_2-k}.\label{BRSaction}
\eea
The Felder complex is built up out of the base cell
\beq\C B_\alpha= \C B^{\B m}_{n_1,n_2}=
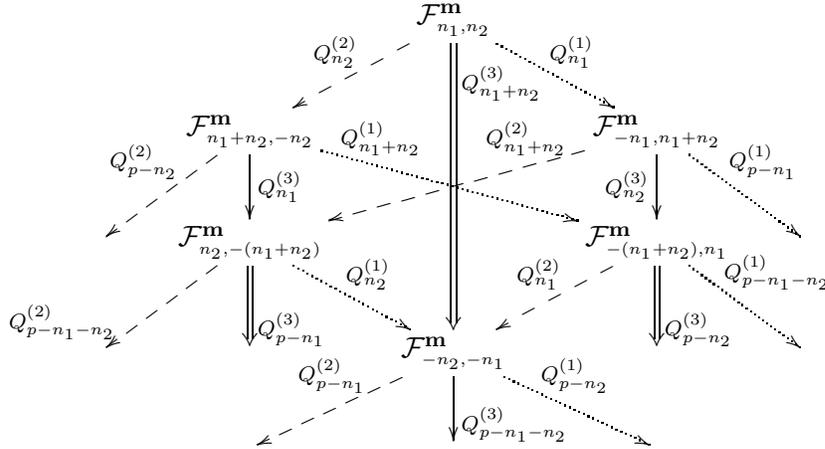
\xymatrix{&&{\C F^{\B m}_{\sst n_1,n_2}}\ar@{.>}[dr]^{Q^{(1)}_{n_1}}\ar@{=>}[ddd]^<<<<<<{Q^{(3)}_{n_1+n_2}}\ar@{-->}[dl]_{Q^{(2)}_{n_2}}&&\\
&{\C F^{\B m}_{\sst n_1+n_2,-n_2}}\ar@{.>}[drr]^<<<<<<<{Q^{(1)}_{n_1+n_2}}\ar[d]^{Q^{(3)}_{n_1}}\ar@{-->}[dl]_{Q^{(2)}_{p-n_2}}& &{\C F^{\B m}_{\sst -n_1,n_1+n_2}}\ar@{-->}[dll]_<<<<<<<{Q^{(2)}_{n_1+n_2}}\ar@{->}[d]_{Q^{(3)}_{n_2}}\ar@{.>}[dr]^{Q^{(1)}_{p-n_1}}& \\
 &          {\C F^{\B m}_{\sst n_2,-(n_1+n_2)}}\ar@{.>}[dr]^{Q^{(1)}_{n_2}} 
\ar@{-->}[dl]_>{Q^{(2)}_{p-n_1-n_2}}\ar@{=>}[d]^>>>{Q^{(3)}_{p-n_1}}
&& \C F^{\B m}_{\sst -(n_1+n_2),n_1}\ar@{-->}[dl]_{Q^{(2)}_{n_1}}\ar@{.>}[dr]^{Q^{(1)}_{p-n_1-n_2}}\ar@{=>}[d]^>>>{Q^{(3)}_{p-n_2}}&\\
  &          &{\C F^{\B m}_{\sst -n_2,-n_1}}\ar@{-->}[dl]_{Q^{(2)}_{p-n_1}}\ar@{->}[d]^>>>{Q^{(3)}_{p-n_1-n_2}}
\ar@{.>}[dr]^{Q^{(1)}_{p-n_2}}&&
\\
&&&\\
}
\label{basecell}\eeq
where maps incoming to the cell have been suppressed.
The operation of the $Q^{(i)}$ on the Fock spaces 
within the base cell 
corresponds to the operation of the Weyl group on the vectors $n_1\omega_1+n_2\omega_2$;
\beq Q^{(i)}\longleftrightarrow  w_i,\eeq
and the operation outside the base cell is the same plus a translation
by $p e_i$.
Note that the action
of $Q^{(3)}$ shown by the double arrow changes the ghost number by 3 and is excluded from the BRST charge defined below.
%
%
\FIGURE[t]{
\xymatrix@=5mm{&{\circ}\ar@{.>}[dr]
\ar@{-->}[dl]&&{\bullet}\ar@{.>}[dr]
\ar@{-->}[dl]&&{\bullet}\ar@{.>}[dr]
\ar@{-->}[dl]&&{\circ}\ar@{.>}[dr]
\ar@{-->}[dl]&&{\bullet}\ar@{.>}[dr]
\ar@{-->}[dl]&&\\
{\bullet}\ar@{.>}[drr]\ar[d]& &{\bullet}\ar@{.>}[drr]\ar@{-->}[dll]\ar[d] & &{\bullet}\ar@{-->}[dll]\ar[d]\ar@{.>}[drr]& &{\bullet}\ar@{-->}[dll]\ar[d]\ar@{.>}[drr]& &{\bullet}\ar@{-->}[dll]\ar[d]\ar@{.>}[drr]& &{\bullet}\ar@{-->}[dll]\ar[d]\\
           {\bullet}\ar@{.>}[dr] && {\bullet}
\ar@{-->}[dl]\ar@{.>}[dr]&& {\circ}
\ar@{-->}[dl]\ar@{.>}[dr]&&{\bullet}
\ar@{-->}[dl]\ar@{.>}[dr]&&{\bullet}
\ar@{-->}[dl]\ar@{.>}[dr]&&{\circ}
\ar@{-->}[dl]\ar@{.>}[dr]\\
&{\bullet}\ar@{.>}[drr]\ar[d]& &{\bullet}\ar@{.>}[drr]\ar@{-->}[dll]\ar[d] & &{\bullet}\ar@{-->}[dll]\ar[d]\ar@{.>}[drr]& &{\bullet}\ar@{-->}[dll]\ar[d]\ar@{.>}[drr]& &{\bullet}\ar@{-->}[dll]\ar[d]\ar@{.>}[drr]& &{\bullet}\ar@{-->}[dll]\ar[d]\\
&{\circ}\ar@{.>}[dr]
\ar@{-->}[dl]&&{\bullet}\ar@{.>}[dr]
\ar@{-->}[dl]&&{\bullet}\ar@{.>}[dr]
\ar@{-->}[dl]&&{\circ}\ar@{.>}[dr]
\ar@{-->}[dl]&&{\bullet}\ar@{.>}[dr]
\ar@{-->}[dl]&&{\bullet}\ar@{-->}[dl]\\
{\bullet}\ar@{.>}[drr]\ar[d]& &{\bullet}\ar@{.>}[drr]\ar@{-->}[dll]\ar[d] & &{\bullet}\ar@{-->}[dll]\ar[d]\ar@{.>}[drr]& &{\bullet}\ar@{-->}[dll]\ar[d]\ar@{.>}[drr]& &{\bullet}\ar@{-->}[dll]\ar[d]\ar@{.>}[drr]& &{\bullet}\ar@{-->}[dll]\ar[d]\\
         {\bullet}\ar@{.>}[dr]&& {\bullet}
\ar@{-->}[dl]\ar@{.>}[dr]&& {\circ}
\ar@{-->}[dl]\ar@{.>}[dr]&&{\bullet}
\ar@{-->}[dl]\ar@{.>}[dr]&&{\bullet}
\ar@{-->}[dl]\ar@{.>}[dr]&&{\circ}
\ar@{-->}[dl]\ar@{.>}[dr]\\
&{\bullet}\ar@{.>}[drr]\ar[d]& &{\bullet}\ar@{.>}[drr]\ar@{-->}[dll]\ar[d] & &{\bullet}\ar@{-->}[dll]\ar[d]\ar@{.>}[drr]& &{\bullet}\ar@{-->}[dll]\ar[d]\ar@{.>}[drr]& &{\bullet}\ar@{-->}[dll]\ar[d]\ar@{.>}[drr]& &{\bullet}\ar@{-->}[dll]\ar[d]\\
&{\circ}\ar@{.>}[dr]&&{\bullet}\ar@{.>}[dr]\ar@{-->}[dl]&&{\bullet}\ar@{.>}[dr]\ar@{-->}[dl]&&{\circ}\ar@{.>}[dr]\ar@{-->}[dl]&&{\bullet}\ar@{.>}[dr]\ar@{-->}[dl]&&{\bullet}\ar@{-->}[dl]\\
   &  &{\bullet}&&{\bullet}&&{\bullet}&&{\bullet}&&{\bullet}\\}
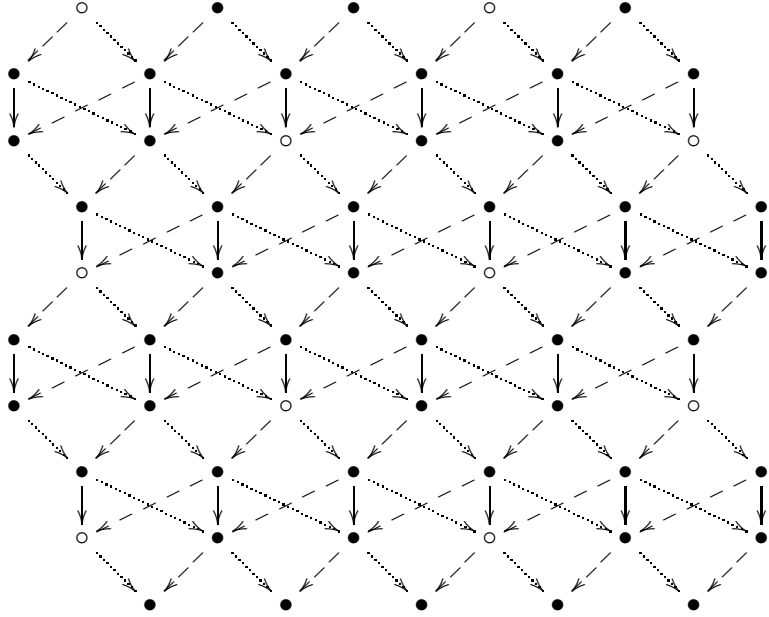
\caption{A part of the complex $\C C_\alpha$. The open circles correspond to the root of the base cell $\C F^{\B m}_{n_1,n_2}$ and its images under translations by
$pe_i$. Different types of arrow have the same meaning as in
\ref{basecell} and only those 
 which are part of $Q_B$ are shown. }
}
%
%
%
%
It is straightforward to convince oneself that repeated application of the
$Q^{(i)}$ generates the infinite complex 
\beq  \C C_\alpha=\bigoplus_{M,N\in\allint} \C B_{\alpha+p\alpha_+(Me_1+Ne_2)},\label{Fcomplex}\eeq
a portion of which is shown in Fig.1. It was shown in \cite{Bouwknegt:1990xa,Mizoguchi:1992vt} that a nillpotent BRST operator $Q_B$
 graded in the $\rho$ direction can be
defined on  $\C C_\alpha$. To do this we first define operators $d_n^{(i)}$ whose action on the Fock spaces is given by $Q_n^{(i)}$ for $i=1,2$, and which increase the ghost number $g$ by one. To keep track of the phases occuring we then define   a cocycle operator $\C N$ with the properties
\bea \C N d_n^{(2)}&=&d_n^{(1)}\C N=0,\quad \C N d_n^{(1)}=d_n^{(1)},\quad d_n^{(2)}=d_n^{(2)}\C N,
\eea
and 
\beq d_n^{(3)}=Q_n^{(3)} (-1)^{1+n\C N},
\eeq
where $[\C N, d^{(3)}]=0$.
The BRST operator $Q_B$ for ghost number $g$ is then given by
\bea Q_B= d^{(1)}+d^{(2)}+d^{(3)}\half(1-(-1)^g),
\eea
and has the property $(Q_B)^2=0$ when acting on $\C C_\alpha  $.
 The Hilbert space $\C H$ is then expected to be  \cite{Felder:1989zp,Bouwknegt:1990xa}
\beq \C H=\frac{\mathrm{Kernel}\, Q_B}{\mathrm{Image}\, Q_B}.\label{coho}\eeq
Therefore $W$ characters for the representations in Table 1 can be calculated by computing the
expectation value of $\mathrm{Tr}\, q^{L_0-c/24}$ from the alternating sum over the BRST complex \cite{Felder:1989zp}
with the result
\begin{equation}
\chi_{ F }(q)=\frac{1}{\eta(\tau)^{2}}
\sum_{  \substack{w\in {\mathcal W}\\ M,N\in {\mathbb Z}}  }\varepsilon_w q^{|p^{\prime}w(\Lambda_{+}(F)%
+\rho)+pp^{\prime}(Ne_{1}+Me_{2})-p(\Lambda_{-}(F)+\rho)|^{2}/2pp'},\label{character}
\end{equation}
where the Dedekind $\eta$ function is given by
\beq \eta(\tau)=q^{1/24}\prod_{k=1}^\infty (1-q^k), \quad q=e^{i2\pi\tau}.\eeq
This agrees with the $W_3$ characters calculated in other ways
(see \cite{Bouwknegt:1993wg} for a review).
 For each one of the representations, this formula gives
the same as for the corresponding
operators in the three state Potts model (see eg \cite{DiFrancesco:1997nk}) which allows us to 
conclude that $W_{3}(4,5)$ does indeed describe the Potts model.

\section{Boundary states in free-field representation}

Coherent boundary states may be defined in a straightforward 
generalization of the procedure for the one component Coulomb gas. First we introduce
the states $|\alpha,\bar\alpha;\alpha_0\rangle$ which are 
 constructed by applying the vertex operator
$V_{\alpha}(z)$ and its antiholomorphic counterpart, ${\overline V}_{\bar\alpha}(\overline{z})$
 to the
$SL(2,C)$-invariant vacuum $|0,0;\alpha_{0}\rangle$,
\begin{equation}
|\alpha,\bar{\alpha};\alpha_{0}
\rangle=\lim_{z,\overline{z}\rightarrow0}\overline V_{\bar\alpha}(\overline
{z})V_{\alpha}(z)|0,0;\alpha_{0}\rangle=e^{i\overline{\alpha}
\cdot\overline \phi_{0}}e^{i\alpha\cdot\phi_{0}}|0,0;\alpha_{0}\rangle.
\end{equation}
These states satisfy
\begin{eqnarray}
 a_{0}^{i}|\alpha,\bar{\alpha}
;\alpha_{0}\rangle&=&\alpha^{i}|\alpha,\bar{\alpha};\alpha_{0}\rangle,\nonumber\\
 \bar{a}_{0}^i|\alpha,\bar{\alpha}
;\alpha_{0}\rangle&=&\bar{\alpha}^{i}|\alpha,\bar{\alpha};\alpha_{0}\rangle.
\end{eqnarray}
The corresponding bra states are given by
\begin{equation}
\langle\alpha,\bar{\alpha};\alpha_{0}|
=\langle 0,0;\alpha_{0}| e^{-i\overline{\alpha}
\cdot\overline \phi_{0}}e^{-i\alpha\cdot\phi_{0}}.
\end{equation}
We then make the  coherent  state ansatz
\bea
|B(\alpha,\bar{\alpha})\rangle&=&C\,|\alpha,\bar{\alpha};\alpha_{0}\rangle,\\
C&=&\prod_{k>0}\exp\left(\frac{1}{k}\,a_{-k}
\cdot\Lambda\cdot \bar{a}_{-k}\right) ,
\eea
where $\Lambda$ is a matrix to be determined by imposing the boundary condition
\begin{equation}
(L_{n}-\bar{L}_{-n})|B(\alpha,\bar{\alpha})\rangle=0.\label{LLbar}
\end{equation}
For positive $n$ we find the constraint
\bea
\Big(&&\half\sum_{l=1}^{n-1}\abar_{n-l}\cdot(\Lambda^T\cdot\Lambda-I)\cdot\abar_{-l}
+(a_0-2\alpha_0(n+1)\rho)\cdot\Lambda\cdot\abar_{-n}\nn\\
&&+(-\abar_0-2\alpha_0(n-1)\rho)\cdot\abar_{-n}\Big) \ket{\alpha,\bar{\alpha};\alpha_{0}}=0,\label{eqn:lambdacond}%
\eea
where $I$ is the identity matrix, and a similar constraint for negative $n$. From these we find the conditions
\bea \Lambda^T\cdot\Lambda&=&I,\nn\\
\Lambda\cdot\rho+\rho&=&0,\nn\\
\Lambda^T\cdot\alpha+4\alpha_0\rho-\bar{\alpha}&=&0.\label{eqn:lambdacondA}
\eea
There are two solutions to (\ref{eqn:lambdacondA}),
\bea C&=&C_{-I}:\quad \Lambda=-I,\quad \alphabar=4\alpha_0\rho-\alpha,\\
C&=&C_{w_\rho}:\quad \Lambda=w_\rho,\quad \alphabar=4\alpha_0\rho+w_\rho \alpha.\label{Csols}
\eea
These  allow us to introduce the abbreviated notation
\begin{equation}
|B(\alpha,\Lambda)\rangle=C_\Lambda \, |\alpha,4\alpha
_{0}\rho+\Lambda\cdot\alpha;\alpha_{0}\rangle.\label{coherentBS}
\end{equation}
 The corresponding out-state is
\begin{equation}
\langle B(\alpha,\Lambda)|= \langle \alpha,4\alpha
_{0}\rho+\Lambda\cdot\alpha;\alpha_{0}|\,C^T_\Lambda,\label{coherentBSout}
\end{equation}
where
\begin{equation}
C^T_\Lambda=\prod_{k>0}\exp\left(\frac{1}{k}a_{k}\cdot\Lambda\cdot\bar{a}_{k}\right) .
\end{equation}

The conservation of the $W$ current across the boundary can be checked
by an explicit calculation.
Since $W(z)$ is a primary field of conformal weight $h=3$, we have
\begin{equation}
[L_{n},W_{m}]=(2n-m)W_{m+n}.
\end{equation}
Putting $m=0$ gives
\begin{equation}
W_{n}=\frac{1}{2n}[L_{n},W_{0}],
\end{equation}
and so
\begin{eqnarray}
(W_{n}+\bar W_{-n})|B(\alpha,\Lambda)\rangle& =&\frac{1}{2n}\left[  (L_{n}-\bar L_{-n})(W_{0}+\bar
W_{0})-(W_{0}+\bar W_{0})(L_{n}-\bar L_{-n})\right] |B(\alpha,\Lambda)\rangle\nonumber\\
&=& \frac{1}{2n}  (L_{n}-\bar L_{-n})(W_{0}+\bar
W_{0})|B(\alpha,\Lambda)\rangle,
\end{eqnarray}
where we have used (\ref{LLbar}).
 After some
calculation using (\ref{Wzero}) we find that 
\begin{equation}
(W_{0}+\bar{W}_{0})|B(\alpha,-I)\rangle=0,
\end{equation}
but that
\begin{equation}
(W_{0}+\bar{W}_{0})|B(\alpha,w_\rho)\rangle\ne0.
\end{equation}
Thus  the $W$ current is conserved for  coherent boundary
states made with the operator $C_{-I}$ but not for those made
with the operator $C_{w_\rho}$.

Using the free-field representation of the boundary states and the
Virasoro operators, we find the cylinder amplitudes
\begin{eqnarray}
\langle B(\alpha,-I)| q^{\Ham} |B(\beta,-I)\rangle
&=&\frac{q^{\half (\beta-2\alpha_{0}\rho)^{2}}}%
{\eta(\tau)^{2}}\delta_{\alpha,{\beta}},\nn\\
\langle B(\alpha,w_\rho)| q^{\Ham} |B(\beta,w_\rho)\rangle
&=&\frac{q^{\half (\beta-2\alpha_{0}\rho)^{2}}}%
{\eta(\tau)^{2}}\delta_{\alpha,{\beta}},\nn\\
\langle B(\alpha,-I)| q^{\Ham} |B(\beta,w_\rho)\rangle
&=&\frac{q^{\half (\beta-2\alpha_{0}\rho)^{2}}}%
{\eta(2\tau)}\delta_{\alpha,{\beta}}\delta_{\alpha,-w_\rho\beta}.\label{cylamp}
\end{eqnarray}
We see immediately that only those representations for which 
 $\alpha$ is proportional to the Weyl vector $\rho$
have non-zero mixed amplitudes between $W$ conserving and $W$ violating boundary states. It is 
straightforward to check by examining Table 1
 that in the  Potts model case  only
$I$ and $\epsilon$ have this property and only for the charges $\alpha$ and 
$4\alpha_0\rho-\alpha$ where
 the  $\alpha$s are drawn from the first column
of Table 1. Thus we can construct  $W$ conserving boundary states 
built on the six operators in the three state Potts model and two extra $W$ violating boundary 
states corresponding to $I$ and $\epsilon$. Linear combinations of these represent the six physical
boundary states first found by Cardy \cite{Cardy:1989ir}  plus the free and 
new boundary states which 
were only found much later \cite{Affleck:1998nq}. It was argued in 
\cite{Fuchs:1998qn} that these eight states are the complete set of boundary states for this model.
In the present formalism  
four other $W$ violating boundary states corresponding to 
$\sigma$, $\sigma^\dagger$, $\psi$ and $\psi^{\dagger}$ can also be written down but 
are completely
decoupled; thus of course they never appear in the fusion rule analysis
of \cite{Fuchs:1998qn}.

The states $\ket {B(\alpha,\Lambda)}$ lie in the Fock spaces. To obtain states in the Hilbert
space of the conformal field theory we must sum over the Felder complex
to obtain 
\bea
\ishiket F&=&\sum_{\substack{w\in \mathcal W\\M,N\in
\mathbb Z}}\kappa^{\phantom\prime}_{wMN}\,
|B(-\alpha
_{+}w^{\ast}(\Lambda_{+}(F))-2\alpha_{0} p'p(N e_{1}+M e_{2})-\alpha_{-}\Lambda
_{-}(F),-I)\rangle,\nn\\
\ishiket\Fbar&=&\sum_{\substack{w\in \mathcal W\\M,N\in
\mathbb Z}}\kappa^{\phantom\prime}_{wMN}\,
|B(-\alpha
_{+}w^{\ast}(\Lambda_{+}(F))-2\alpha_{0} p'p(N e_{1}+M e_{2})-\alpha_{-}\Lambda
_{-}(F),w_\rho)\rangle,\nn\\ &&\label{notIshi}
\eea
where the constants $\kappa^{\phantom\prime}_{wMN}$ have magnitude 1.
There are similar expressions for the bra states but with $\kappa^{\phantom\prime}_{wMN}$ replaced
by $\kappa_{wMN}'$ satisfying
\beq \kappa^{\phantom\prime}_{wMN}\kappa_{wMN}'=\varepsilon_w.\eeq
Using (\ref{character})  and (\ref{cylamp}),
we find the cylinder amplitudes
\bea \ishibra {F'} q^{\Ham}\ishiket F &=&\chi_{F}(q)\delta_{F,F'},\nn\\
\ishibra {\Fbar'} q^{\Ham}\ishiket \Fbar &=&\chi_{F}(q)\delta_{F,F'},\nn\\
\ishibra \Ibar q^{\Ham}\ishiket I &=& \frac{q^{-\frac{1}{30}}}{\prod_{k>0}(1-q^{2k})}\sum_{n\in\allint}q^{20n^2+2n}-q^{20n^2+18n+4},\nn\\
\ishibra \Ebar q^{\Ham} \ishiket \epsilon &=& \frac{q^{\frac{11}{30}}}{\prod_{k>0}(1-q^{2k})}\sum_{n\in\allint}q^{20n^2+6n}-q^{20n^2+14n+2},\label{Wamps}
\eea
with all other mixed amplitudes being zero.

We can compare these results with the discussion of Affleck et al \cite{Affleck:1998nq}. 
The $c=4/5$ models are Virasoro theories with $(p,q)=(5,6)$  and on
each conformal tower $(r,s)$ can be built 
 \emph{Virasoro} Ishibashi states $\vert r,s \rangle\!\rangle$ which are related
to the characters by
\beq \chi_{r,s}=\langle\!\langle r,s\vert  q^{\Ham} 
\vert r,s \rangle\!\rangle.\eeq
 Following \cite{Affleck:1998nq}  we expect
that the cylinder amplitudes for the Potts model 
should be given in terms of the 
Virasoro characters by
\bea\ishibra I q^{\Ham}\ishiket I=\ishibra \Ibar q^{\Ham}\ishiket \Ibar
&=&\chi_{1,1}+\chi_{4,1},\nn\\
\ishibra \Ibar q^{\Ham}\ishiket I
&=&\chi_{1,1}-\chi_{4,1},\nn\\ 
\ishibra \epsilon q^{\Ham}\ishiket \epsilon=\ishibra \Ebar q^{\Ham}\ishiket 
\Ebar
&=&\chi_{2,1}+\chi_{3,1},\nn\\
\ishibra \Ebar q^{\Ham}\ishiket \epsilon
&=&\chi_{2,1}-\chi_{3,1},\nn\\ 
\ishibra \sigma q^{\Ham}\ishiket \sigma=\ishibra{\sigma^\dagger}  q^{\Ham}\ishiket {\sigma^\dagger}
&=&\chi_{3,3},\nn\\
\ishibra \psi q^{\Ham}\ishiket \psi=\ishibra{\psi^\dagger}  q^{\Ham}\ishiket {\psi^\dagger}
&=&\chi_{4,3}.\label{WVir}\eea
It can be checked that the identities between Riemann-Jacobi functions implied by these
relationships and our results (\ref{Wamps}) are all true; as an example
we give a proof in Appendix B for a case which has not appeared in the literature before,
namely the mixed amplitude for the identity operator.

In terms of these Ishibashi states Cardy's
 physical boundary states \cite{Cardy:1989ir}
take the usual form for the three state Potts model, 
\bea
 |\tilde I\rangle&=&N\left\{ (|I\rangle\!\rangle+|\psi\rangle\!\rangle
+|\psi^{\dag}\rangle\!\rangle)+\lambda(|\epsilon\rangle\!\rangle
+|\sigma\rangle\!\rangle+|\sigma^{\dag}\rangle\!\rangle)\right\} ,\nn\\
 |\tilde\epsilon\rangle&=&N\left\{ \lambda^{2}(|I\rangle\!\rangle+|\psi
\rangle\!\rangle+|\psi^{\dag}\rangle\!\rangle)-\lambda^{-1}(|\epsilon
\rangle\!\rangle+|\sigma\rangle\!\rangle+|\sigma^{\dag}\rangle\!\rangle
)\right\} ,\nn\\
 |\tilde\psi\rangle&=&N\left\{ (|I\rangle\!\rangle+\omega|\psi\rangle
\!\rangle+\bar\omega|\psi^{\dag}\rangle\!\rangle)+\lambda(|\epsilon
\rangle\!\rangle+\omega|\sigma\rangle\!\rangle+\bar\omega|\sigma^{\dag}%
\rangle\!\rangle)\right\} ,\nn\\
 |\tilde\psi^{\dag}\rangle&=&N\left\{ (|I\rangle\!\rangle+\bar\omega
|\psi\rangle\!\rangle+\omega|\psi^{\dag}\rangle\!\rangle)+\lambda
(|\epsilon\rangle\!\rangle+\bar\omega|\sigma\rangle\!\rangle+\omega
|\sigma^{\dag}\rangle\!\rangle)\right\} ,\nn\\
 |\tilde\sigma\rangle&=&N\left\{ \lambda^{2}(|I\rangle\!\rangle+\omega
|\psi\rangle\!\rangle+\bar\omega|\psi^{\dag}\rangle\!\rangle)-\lambda
^{-1}(|\epsilon\rangle\!\rangle+\omega|\sigma\rangle\!\rangle+\bar
\omega|\sigma^{\dag}\rangle\!\rangle)\right\} ,\nn\\
 |\tilde\sigma^{\dag}\rangle&=&N\left\{ \lambda^{2}(|I\rangle\!\rangle
+\bar\omega|\psi\rangle\!\rangle+\omega|\psi^{\dag}\rangle\!\rangle
)-\lambda^{-1}(|\epsilon\rangle\!\rangle+\bar\omega|\sigma\rangle
\!\rangle+\omega|\sigma^{\dag}\rangle\!\rangle)\right\} ,\label{idstate}
\eea
where
\begin{equation}
N=\sqrt{\frac{2}{\sqrt{15}}\sin\frac{\pi}{5}},\;\;\;\;\; \lambda=\sqrt
{\frac{\sin(2\pi/5)}{\sin(\pi/5)}},
\end{equation}
and
\begin{equation}
\omega=e^{2\pi i/3},\;\;\;\;\; \bar\omega=e^{4\pi i/3}.
\end{equation}
These states represent the fixed and mixed boundary conditions for the Potts model spins. In addition there are two more physical boundary states
\bea \ket {\mathrm {free}} &=& N\sqrt{3}\left(\ishiket\Ibar-\lambda\ishiket\Ebar\right),\nn\\
\ket{\mathrm {new}} &=& N\sqrt{3}\left(\lambda^2\ishiket\Ibar+\lambda^{-1}\ishiket\Ebar\right),\label{freenew}
\eea
which represent the free and so-called new boundary conditions for the 
spin model.

\section{Critical Potts model correlation functions with boundary}

We will now show how the free-field formalism can be used to compute exact correlation functions of 
bulk operators in the presence of a boundary when the boundary condition
 is represented by the  states (\ref{idstate}, \ref{freenew}).

\subsection{General considerations}

We consider a unit disk in the $z$-plane,  where a boundary with condition $\tilde\alpha$ is 
placed at $|z|=1$.
At the origin we place the M\"obius invariant vacuum $\vert 0,0;\alpha_0\rangle$.
Then  $p$-point correlation functions on the disk are given by
\begin{equation}
\langle\tilde\alpha\vert
V_{\alpha_1}(z_1)\bar V_{\bar\alpha_1}(\bar z_1)\cdots
V_{\alpha_p}(z_p)\bar V_{\bar\alpha_p}(\bar z_p)
\times (\mbox{screening operators})
\vert 0,0;\alpha_0\rangle.
\label{eqn:correlator1}
\end{equation}
As a Cardy state $\langle\tilde\alpha\vert$ is a linear sum of the Ishibashi states 
$\lishi F\vert$ and $\lishi\Fbar \vert$ which are in turn linear sums of the coherent states 
$\langle B(\alpha, \Lambda)\vert$, 
finding the correlation functions (\ref{eqn:correlator1}) boils down to evaluating the disk amplitudes with boundary charges $\alpha$ and boundary types $\Lambda$,
\bea
&&\langle B(\alpha,\Lambda)\vert
V_{\alpha_1}(z_1)\bar V_{\bar\alpha_1}(\bar z_1)\cdots
V_{\alpha_p}(z_p)\bar V_{\bar\alpha_p}(\bar z_p)\nonumber\\
&&\times(Q_+^{(1)})^{m_1} (Q_-^{(1)})^{n_1} (Q_+^{(2)})^{m_2} (Q_-^{(2)})^{n_2} 
(\bar Q_+^{(1)})^{\bar m_{1}}(\bar Q_-^{(1)})^{\bar n_{1}}
(\bar Q_+^{(2)})^{\bar m_{2}}(\bar Q_-^{(2)})^{\bar n_{2}}
\vert 0,0;\alpha_0\rangle,
\label{eqn:correlator2}
\eea
where $m_1$, etc. are numbers of the screening operators.
Physical correlation functions are simply linear combinations of the amplitudes (\ref{eqn:correlator2}).
Since the inner products of the Fock spaces are non-vanishing only when both the holomorphic and antiholomorphic charge neutrality conditions are satisfied, the amplitudes (\ref{eqn:correlator2}) are subject to the conditions,
\bea
&&-\alpha+\alpha_1+\cdots+\alpha_p
+m_1\alpha_+e_1+m_2\alpha_+e_2+n_1\alpha_-e_1+n_2\alpha_-e_2=0,\\
&&-4\alpha_0\rho-\Lambda\cdot\alpha+\bar\alpha_1+\cdots+\bar\alpha_p
+\bar m_1\alpha_+e_1+\bar m_2\alpha_+e_2+\bar n_1\alpha_-e_1+\bar n_2\alpha_-e_2=0,
\eea
otherwise they vanish.
These neutrality conditions relate the boundary charges $\alpha$ to possible configurations of the screening operators.
The disk amplitudes are evaluated using the formula,
\bea\lefteqn{\langle B(\alpha, \Lambda)\vert
\prod_{i=1}^M V_{\alpha_i}(z_i) \prod_{j=1}^N \bar V_{\alphabar_j}(\bar z_j)\vert 0,0;\alpha_0\rangle
=\delta_{\alpha,\sum\alpha_i}
\delta_{4\alpha_0\rho+\Lambda\cdot\alpha,\sum\alphabar_j}}\nonumber\\
&&\times\left\{\prod_{i<j}^M (z_i-z_j)^{\alpha_i\cdot\alpha_j}\right\}
\left\{\prod_{i<j}^N (\bar z_i-\bar z_j)^{\alphabar_i\cdot\alphabar_j}\right\}
\left\{\prod_{i=1}^M\prod_{j=1}^N(1-z_i\bar z_j)^{-\alpha_i\cdot\Lambda\cdot\alphabar_j}\right\}
.\nn\\
\eea
The presence of  screening operators leads to integral representations for the amplitudes.
The integration contours are originally Felder's contours (concentric circles around the 
origin and attached to vertex operators). 
It is argued \cite{Kawai:2002pz} that at least in the cases of 1 and 2-point functions these
 contours can be deformed suitably to give convergent functions which correspond to certain conformal blocks.  
We will now use  this simple setup  to  compute Potts model correlation functions.

\subsection{Vertex operators}

In order to compute disk correlation functions in the Coulomb gas formalism, we need to define non-chiral primary operators in terms of chiral operators and then represent them with vertex operators.

The first step involves some intricacies because the non-chiral spin operators for instance can be either 
\beq
\sigma(z,\bar z)=\sigma(z)\otimes\bar\sigma(\bar z),\;\;
\sigma^\dag(z,\bar z)=\sigma^\dag(z)\otimes\bar\sigma^\dag(\bar z),
\label{eqn:NonChiralSpin1}
\eeq
or 
\beq
\sigma(z,\bar z)=\sigma(z)\otimes\bar\sigma^\dag(\bar z),\;\;
\sigma^\dag(z,\bar z)=\sigma^\dag(z)\otimes\bar\sigma(\bar z).
\label{eqn:NonChiralSpin2}
\eeq
Although the first one (\ref{eqn:NonChiralSpin1}) might seem more natural, this does not represent the Potts model correctly since it would lead to vanishing spin 1-point functions even near a fixed boundary.
Thus we shall adopt the definition (\ref{eqn:NonChiralSpin2}) for the non-chiral spin operators and similarly we define
\beq
\psi(z,\bar z)=\psi(z)\otimes\bar\psi^\dag(\bar z),\;\;
\psi^\dag(z,\bar z)=\psi^\dag(z)\otimes\bar\psi(\bar z).
\label{eqn:NonChiralPsi}
\eeq
The identity and energy operators do not have this ambiguity and we shall simply define
\beq
I(z,\bar z)=I(z)\otimes\bar I(\bar z),\;\;
\epsilon(z,\bar z)=\epsilon(z)\otimes\bar\epsilon(\bar z).
\label{eqn:NonChiralIE}
\eeq

In the second step, we need to represent each $W$ primary field by a vertex operator having one of the six charges of (\ref{primarydefn}).
In the bulk theory and also in presence of a $W$ conserving boundary, one may allocate whichever of the six charges
is most convenient  as they should be all equivalent under the chiral algebra.
However, as we have seen, in the presence of $W$ violating boundary conditions
this equivalence breaks down 
and only charges associated with 
 the first column of Table 1 are allowed.
We are still free to use $\alpha$ or its conjugate $4\alpha_0\rho+w_\rho\alpha$ to construct a chiral vertex operator
and  can, in particular, represent a non-chiral operator as a product of different holomorphic and antiholomorphic vertex operators, e.g. 
$I(z,\bar z)=V_{0}(z)\otimes\bar V_{4\alpha_0\rho}(\bar z)$.
In practice we shall choose a set of vertex operators which minimizes the number of  screening operators. 

\subsection{One point functions} 

In the case of the spin 1-point function $\langle\sigma(z,\bar z)\rangle_{boundary}$ with various
boundary conditions represented by the Cardy states,
the number of  screening operators is minimized by defining
$\sigma(z,\bar z)=V_{\alpha_1}(z)\otimes\bar V_{4\alpha_0\rho-\alpha_1}(\bar z)$,
where $\alpha_1$ is either $\alpha_1=-\alpha_-\omega_1$ or its conjugate
$4\alpha_0\rho+\alpha_-\omega_2$.
Then the 1-point function is a linear sum of the disk 1-point amplitudes,
\bea
&&\langle B(\alpha,\Lambda)\vert
V_{\alpha_1}(z_1)\bar V_{4\alpha_0\rho-\alpha_1}(\bar z_1)
(Q_+^{(1)})^{m_1} (Q_-^{(1)})^{n_1}\nonumber\\
&&\times (Q_+^{(2)})^{m_2} (Q_-^{(2)})^{n_2}
(\bar Q_+^{(1)})^{\bar m_{1}}(\bar Q_-^{(1)})^{\bar n_{1}}
(\bar Q_+^{(2)})^{\bar m_{2}}(\bar Q_-^{(2)})^{\bar n_{2}}
\vert 0,0;\alpha_0\rangle,
\eea
 subject to the charge neutrality conditions
\bea
&&-\alpha+\alpha_1
+m_1\alpha_+e_1+m_2\alpha_+e_2+n_1\alpha_-e_1+n_2\alpha_-e_2=0,
\label{eqn:1pNeutrality1}\\
&&-\Lambda\cdot\alpha-\alpha_1
+\bar m_1\alpha_+e_1+\bar m_2\alpha_+e_2+\bar n_1\alpha_-e_1+\bar n_2\alpha_-e_2=0.
\label{eqn:1tNeutrality2}
\eea
Adding the above two expressions we have
\beq
(I+\Lambda)\cdot\alpha=
(m_1+\bar m_1)\alpha_+e_1+(m_2+\bar m_2)\alpha_+e_2
+(n_1+\bar n_1)\alpha_-e_1+(n_2+\bar n_2)\alpha_-e_2.
\label{eqn:1pNeutrality3}
\eeq
For $\Lambda=-I$, this condition is satisfied by
$m_1=\bar m_1=m_2=\bar m_2=n_1=\bar n_1=n_2=\bar n_2=0$, that is, with no screening operators, and from (\ref{eqn:1pNeutrality1}) we have 
$\alpha=\alpha_1$.
Thus in this case, 
\begin{equation}
\langle B(\alpha, -I)\vert V_{\alpha_1}(z)\bar V_{4\alpha_0\rho-\alpha_1}(\bar z)\vert 0,0;\alpha_0\rangle
=(1-z\bar z)^{-2h}\delta_{\alpha,\alpha_1},
\label{eqn:1pAmp}
\end{equation}
where $h=\frac 12 \alpha_1 \cdot (\alpha_1-4\alpha_0\rho)=1/15$,
is the only non-trivial disk amplitude.
This means that the charge neutrality condition picks up the coefficient of 
$\langle B(\alpha_1, -I)\vert$ from the boundary state.
When $\Lambda=w_\rho$, the left hand side of (\ref{eqn:1pNeutrality3}) is proportional to $e_1-e_2$ and it cannot be neutralized with non-negative powers of the screening operators. 
The only exception is when $(I+w_\rho)\alpha=0$, but this does not happen in the case of the spin 1-point function.
Thus the charge neutrality can never be satisfied for $\Lambda=w_\rho$ and we have no contribution from $\langle B(\alpha, w_\rho)\vert$ boundary states.

Thus, for example, the spin 1-point function with the boundary condition $\tilde I$ will be
(after normalization)
\beq
\frac{\langle\tilde I\vert V_{\alpha_1}(z)\bar V_{4\alpha_0\rho-\alpha_1}(\bar z)\vert 0,0;\alpha_0\rangle}{\langle\tilde I\vert 0,0;\alpha_0\rangle}
=\lambda(1-z\bar z)^{-2/15}.
\eeq
This 1-point function on the disk is conveniently mapped onto the half-plane, via
the global conformal transformation,
\begin{equation}
w=-iy_0\frac{z-1}{z+1},\;\;\;
\bar w=iy_0\frac{\bar z-1}{\bar z+1},
\label{eqn:conformaltr}
\end{equation}
which takes the origin $(z, \bar z)=(0,0)$ to $(w,\bar w)=(iy_0,-iy_0)$.
The spin 1-point function on the half-plane is then,
\beq
\langle\sigma\rangle_{\tilde I}=\lambda(2y)^{-2/15},
\eeq
where $y$ is the distance from the boundary.
Similarly the spin 1-point function with other boundary conditions are (on the half-plane),
\bea
&&\langle\sigma\rangle_{\tilde\psi}
=\omega\lambda(2y)^{-2/15},\\
&&\langle\sigma\rangle_{\tilde\psi^\dag}
=\bar\omega\lambda(2y)^{-2/15},\\
&&\langle\sigma\rangle_{\tilde\epsilon}
=-\frac{1}{\lambda^3}(2y)^{-2/15},\\
&&\langle\sigma\rangle_{\tilde\sigma}
=-\frac{\omega}{\lambda^3}(2y)^{-2/15},\\
&&\langle\sigma\rangle_{\tilde\sigma^\dag}
=-\frac{\bar\omega}{\lambda^3}(2y)^{-2/15}\\
&&\langle\sigma\rangle_{\mathrm{free}}
=0,\\
&&\langle\sigma\rangle_{\mathrm{new}}
=0.
\eea
The first six Cardy states are identified as the fixed and mixed boundary conditions,
\bea
{\rm Fixed:}&& \tilde I = (A),\;\; \tilde\psi = (B),\;\; \tilde\psi^\dag = (C),\\
{\rm Mixed:}&& \tilde\epsilon = (BC),\;\; \tilde\sigma = (CA),\;\; \tilde\sigma^\dag = (AB).
\eea
The free and new boundary conditions are not immediately distinguishable from the spin 1-point function only.

Next we consider the energy 1-point function. 
For definiteness we let
$\epsilon(z,\bar z)=V_{\alpha_1}(z)\otimes\bar V_{4\alpha_0\rho-\alpha_1}(\bar z)$,
with
$\alpha_1=-\alpha_-(\omega_1+\omega_2)$.
The computation is almost the same as above.
For $\Lambda=-I$ boundary, the charge neutrality condition is satisfied with no screening operators, and (\ref{eqn:1pAmp}) with $h=2/5$ is the only non-trivial contribution in this case.
However, the charge neutrality condition is fulfilled for $\Lambda=w_\rho$ as well, with
$\alpha=\alpha_1$ and no screening operators. 
Thus the amplitude 
\begin{equation}
\langle B(\alpha, w_\rho)\vert V_{\alpha_1}(z)\bar V_{4\alpha_0\rho-\alpha_1}(\bar z)\vert 0,0;\alpha_0\rangle
=(1-z\bar z)^{-4/5}\delta_{\alpha,\alpha_1}
\end{equation}
also contributes in this case.
The resulting energy 1-point function for various boundary conditions are (on the half plane)
\bea
&&\langle\epsilon\rangle_{\tilde I}
=\langle\epsilon\rangle_{\tilde\psi}
=\langle\epsilon\rangle_{\tilde\psi^\dag}
=\lambda(2y)^{-4/5},\\
&&\langle\epsilon\rangle_{\tilde\epsilon}
=\langle\epsilon\rangle_{\tilde\sigma}
=\langle\epsilon\rangle_{\tilde\sigma^\dag}
=-\frac{1}{\lambda^3}(2y)^{-4/5},\\
&&\langle\epsilon\rangle_{\mathrm{free}}=-\lambda(2y)^{-4/5},\\
&&\langle\epsilon\rangle_{\mathrm{new}}=\frac{1}{\lambda^3}(2y)^{-4/5}.
\eea

As we have seen, the 1-point functions come
 along with the coefficients of the corresponding Ishibashi states in the Cardy states,
 and this is in agreement with the commonly accepted  understanding\footnote{In our notation, which is 
traditional in the literature,
 the bra-Ishibashi state $\lishi\sigma\vert$ couples to $\sigma$ but
 it is $\sigma^\dag$ which couples to the ket-Ishibashi state $\vert\sigma\rishi$.
In this sense it would be more natural to exchange the definitions of $\vert\sigma\rishi$ and
$\vert\sigma^\dag\rishi$.} 
 that such coefficients are essentially the 1-point bulk-brane coupling constants \cite{Cardy:1991tv}.

\subsection{Spin two point functions}
Next, we consider the spin 2-point functions 
\beq \langle\sigma(z_1,\bar z_1)\sigma(z_2,\bar z_2)\rangle_{boundary}\eeq
 and
\beq \langle\sigma^\dag(z_1,\bar z_1)\sigma(z_2,\bar z_2)\rangle_{boundary}\eeq
and find their exact forms for various boundary conditions.
In principle they should also be obtainable by the conventional approach \cite{Cardy:1984bb};
 that is, by solving a 6th order differential equation arising from the singular vector at level 6,
\bea
&&\left(
{L}_{-6} -\frac{96367}{136856}{L}_{-1}{L}_{-5} 
-\frac{1630}{17107}{L}_{-2}{L}_{-4} 
+\frac{33795}{136856}{L}_{-1}^2{L}_{-4} -\frac{2437}{102642}{L}_{-3}^2 \right.\nonumber\\
&&+\frac{4909}{68428}{L}_{-1}{L}_{-2}{L}_{-3}-\frac{16905}{273712}{L}_{-1}^3{L}_{-3}
+\frac{576}{85535}{L}_{-2}^3-\frac{651}{34214}{L}_{-1}^2{L}_{-2}^2 \nonumber\\
&&\left.+\frac{3465}{273712}{L}_{-1}^4{L}_{-2}
-\frac{675}{547424}{L}_{-1}^6
\right)\vert\phi_{2,3}\rangle,
\label{eqn:singvec6}
\eea
and then fixing their coefficients using the solutions of sewing relations.
The Coulomb gas calculation presented below is equivalent, but practically it is considerably simpler.

Let us consider $\langle\sigma\sigma\rangle_{boundary}$ first.
We represent the primary operators,
$\sigma(z_1,\bar z_1)=V_1(z_1)\otimes\bar V_1(\bar z_1)$, 
$\sigma(z_2,\bar z_2)=V_2(z_2)\otimes\bar V_2(\bar z_2)$,
by the vertex operators having charges,
\bea
&& \alpha_1=-\alpha_-\omega_1,\\
&& \alpha_2=-\alpha_-\omega_1,\\
&& \bar\alpha_1=-\alpha_-\omega_2,\\
&& \bar\alpha_2=4\alpha_0\rho+\alpha_-\omega_1.
\eea
The holomorphic and antiholomorphic charge neutrality conditions on the disk are,
\bea
&&-\alpha-2\alpha_-\omega_1+m_1\alpha_+e_1
+m_2\alpha_+e_2+n_1\alpha_-e_1+n_2\alpha_-e_2=0,
\label{eqn:sigmacharge1}\\
&&-\Lambda\cdot\alpha+\alpha_-(\omega_1-\omega_2)
+\bar m_1\alpha_+e_1+\bar m_2\alpha_+e_2+\bar n_1\alpha_-e_1+\bar n_2\alpha_-e_2=0.
\label{eqn:sigmacharge2}
\eea
Summing (\ref{eqn:sigmacharge1}) and (\ref{eqn:sigmacharge2}), we have
\bea (m_1+\bar m_1)\alpha_+e_1+(m_2+\bar m_2)\alpha_+e_2
&+&(n_1+\bar n_1)\alpha_-e_1+(n_2+\bar n_2)\alpha_-e_2\nonumber\\
&&=(I+\Lambda)\cdot\alpha+\alpha_-(\omega_1+\omega_2)\nonumber\\
&&=(I+\Lambda)\cdot\alpha+\alpha_-(e_1+e_2).
\label{eqn:2pNeutrality1}
\eea
It can be shown that when $\Lambda=w_\rho$ the charge neutrality is not satisfied. 
For $\Lambda=-I$, (\ref{eqn:2pNeutrality1}) implies
\begin{equation}
m_1=\bar m_1=m_2=\bar m_2=0,\;\;n_1+\bar n_1=1,\;\;n_2+\bar n_2=1.
\end{equation}
There are potentially four cases where the charge neutrality is satisfied,
\bea
({\rm I}): &n_1=n_2=1,\;\;\bar n_1=\bar n_2=0,&\alpha=\alpha_-(\omega_2-\omega_1),\\
({\rm II}): &n_1=\bar n_2=1,\;\;\bar n_1=n_2=0,&\alpha=-\alpha_-\omega_2,\\
({\rm III}): &\bar n_1=n_2=1,\;\; n_1=\bar n_2=0,&\alpha=\alpha_-(2\omega_2-3\omega_1),\\
({\rm IV}): &\bar n_1=\bar n_2=1,\;\; n_1=n_2=0,&\alpha=-2\alpha_-\omega_1.
\eea
We notice that the charge configurations (I) and (III) can never be realized since the corresponding boundary charges do not exist in the Cardy states.
The boundary charges of (II) and (IV) are included in the Ishibashi states $\lishi\sigma^\dag\vert$ and $\lishi\psi^\dag\vert$ respectively, indicating which intermediate state is reflected by the boundary. 
This of course is in agreement with the fusion rules,
\bea
&&\sigma\times\sigma=\psi^\dag+\sigma^\dag,\\
&&\sigma^\dag\times\sigma^\dag=\psi+\sigma.
\eea
We shall denote the disc 2-point amplitudes corresponding to (II) and (IV) as $I_{\sigma}$ and $I_{\psi}$, respectively.
They correspond to the two conformal blocks depicted in the figure below. 

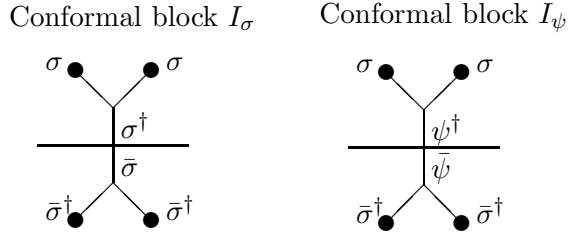
\setlength{\unitlength}{1mm}
\begin{center}
\begin{minipage}{40mm}
Conformal block $I_{\sigma}$

~~~\begin{picture}(20,30)
{\thicklines
\put(0,15){\line(1,0){20}}}
{\thinlines
\put(10,20){\line(0,-1){10}}
\put(10,20){\line(1,1){5}}
\put(10,20){\line(-1,1){5}}
\put(10,10){\line(1,-1){5}}
\put(10,10){\line(-1,-1){5}}}
\put(5,25){\circle*{2}}
\put(15,25){\circle*{2}}
\put(5,5){\circle*{2}}
\put(15,5){\circle*{2}}
\put(1,25){$\sigma$}
\put(17,25){$\sigma$}
\put(1,5){$\bar\sigma^\dag$}
\put(17,5){$\bar\sigma^\dag$}
\put(11,16){$\sigma^\dag$}
\put(11,11){$\bar\sigma$}
\end{picture}
\end{minipage}
\begin{minipage}{40mm}
Conformal block $I_{\psi}$

~~~\begin{picture}(20,30)
{\thicklines
\put(0,15){\line(1,0){20}}}
{\thinlines
\put(10,20){\line(0,-1){10}}
\put(10,20){\line(1,1){5}}
\put(10,20){\line(-1,1){5}}
\put(10,10){\line(1,-1){5}}
\put(10,10){\line(-1,-1){5}}}
\put(5,25){\circle*{2}}
\put(15,25){\circle*{2}}
\put(5,5){\circle*{2}}
\put(15,5){\circle*{2}}
\put(1,25){$\sigma$}
\put(17,25){$\sigma$}
\put(1,5){$\bar\sigma^\dag$}
\put(17,5){$\bar\sigma^\dag$}
\put(11,16){$\psi^\dag$}
\put(11,11){$\bar\psi$}
\end{picture}
\end{minipage}
\end{center}

The evaluation of the amplitudes $I_{\sigma}$ and $I_{\psi}$ is straightforward.
As (II) involves $Q_-^{(1)}$ in the holomorphic and $\bar Q_-^{(2)}$ in the antiholomorphic sector,
we have
\bea
I_{\sigma}&=&\langle B(-\alpha_-\omega_2, -I)\vert V_1(z_1)\bar V_1(\bar z_1)
Q_-^{(1)}\bar Q_-^{(2)}V_2(0)\bar V_2(0)\vert 0,0;\alpha_0\rangle\nonumber\\
&=&
z_1^{8/15}\bar z_1^{2/15}(1-z_1\bar z_1)^{4/15}
\oint du\oint d\bar v 
(z_1-u)^{-4/5}u^{-4/5}(\bar z_1-\bar v)^{-4/5}
\bar v^{-2/5}(1-u\bar v)^{-4/5}.\nonumber\\
\eea
We have used the global conformal invariance to set $z_2=\bar z_2=0$
 (this is equivalent to sending one of the four points to zero and another to infinity in the corresponding chiral four point function).
As the screening operator $Q_-^{(1)}$ lies in the holomorphic sector, the convergent integral must be
proportional to the $u$-integral between $z_1$ and $z_2$.  
Similarly, the antiholomorphic screening operator $\bar Q_-^{(2)}$ leads to $\bar v$-integration
between $\bar z_1$ and $\bar z_2$. 
Hence the integration contours may be deformed into
\begin{equation}
\oint du\oint d\bar v\rightarrow \int_{z_2}^{z_1}du\int_{\bar z_2}^{\bar z_1}d\bar v,
\label{eqn:condef1}
\end{equation}
and we have
\beq
I_{\sigma}=N_{\sigma}\xi^{-1/15}(1-\xi)^{4/15}F(\frac 15, \frac 35, \frac 25;\xi),
\label{eqn:Isigma}
\eeq
where $\xi=z_1\bar z_1$. 
The normalization of the conformal block is fixed by its off-boundary limit and the constant $N_{\sigma}$ is identified with the 3-point coupling constant,
\beq
N_{\sigma}=
C_{\sigma\sigma}{}^{\sigma^\dag}=
\frac{\Gamma(1/5)\Gamma(3/5)}{\Gamma(2/5)^2\lambda},
\eeq
appearing in the bulk OPE \cite{Mizoguchi:1991pf},
\beq
\sigma(z_1,\bar z_1)\sigma(z_2,\bar z_2)
=C_{\sigma\sigma}{}^{\sigma^\dag}|z_1-z_2|^{-2/15}\sigma^\dag(z_2,\bar z_2) 
+C_{\sigma\sigma}{}^{\psi^\dag} |z_1-z_2|^{16/15}\psi^\dag(z_2,\bar z_2)+\cdots.
\eeq
The other conformal block $I_{\psi}$ is similarly evaluated as
\bea
I_{\psi}
&=&\langle B(-2\alpha_-\omega_1,-I)
\vert V_1(z_1)\bar V_1(\bar z_1)\bar Q_-^{(1)}\bar Q_-^{(2)}V_2(0)\bar V_2(0)\vert 0,0;\alpha_0\rangle\nonumber\\
&=&N_{\psi}\xi^{8/15}(1-\xi)^{4/15}F(\frac 45, \frac 65, \frac 85;\xi),
\label{eqn:Ipsi}
\eea
where 
\beq
N_{\psi}=
C_{\sigma\sigma}{}^{\psi^\dag}=
1/3.
\eeq
The spin 2-point function $\langle\sigma\sigma\rangle_{boundary}$
with various boundary conditions is then given by
 linear combinations of $I_{\sigma}$ and $I_{\psi}$, with the
 coefficients of the Ishibashi states $\lishi\sigma^\dag\vert$ and 
$\lishi\psi^\dag\vert$ picked out from the Cardy states.
On the disk we find
\bea
(A):&&\frac{\langle\tilde I\vert\sigma(z_1,\bar z_1)\sigma(0,0)\vert 0,0;\alpha_0\rangle}
{\langle\tilde I\vert 0,0;\alpha_0\rangle}
=I_{\psi}+\lambda I_{\sigma},\\
(B):&&\frac{\langle\tilde\psi\vert\sigma(z_1,\bar z_1)\sigma(0,0)\vert 0,0;\alpha_0\rangle} 
{\langle\tilde\psi\vert 0,0;\alpha_0\rangle}
=\bar\omega(I_{\psi}+\lambda I_{\sigma}),\\
(C):&&\frac{\langle\tilde\psi^\dag\vert\sigma(z_1,\bar z_1)\sigma(0,0)\vert 0,0;\alpha_0\rangle}
{\langle\tilde\psi^\dag\vert 0,0;\alpha_0\rangle}
=\omega(I_{\psi}+\lambda I_{\sigma}),\\
(BC):&&\frac{\langle\tilde\epsilon\vert\sigma(z_1,\bar z_1)\sigma(0,0)\vert 0,0;\alpha_0\rangle} 
{\langle\tilde\epsilon\vert 0,0;\alpha_0\rangle}
=I_{\psi}-\frac{1}{\lambda^3}I_{\sigma},\\
(CA):&&\frac{\langle\tilde\sigma\vert\sigma(z_1,\bar z_1)\sigma(0,0)\vert 0,0;\alpha_0\rangle} 
{\langle\tilde\sigma\vert 0,0;\alpha_0\rangle}
=\bar\omega(I_{\psi}-\frac{1}{\lambda^3}I_{\sigma}),\\
(AB):&&\frac{\langle\tilde\sigma^\dag\vert\sigma(z_1,\bar z_1)\sigma(0,0)
\vert 0,0;\alpha_0\rangle} {\langle\tilde\sigma^\dag\vert 0,0,\alpha_0\rangle}
=\omega(I_{\psi}-\frac{1}{\lambda^3}I_{\sigma}),\\
(\mathrm{free}):&&\frac{\langle \mathrm{free}\vert\sigma(z_1,\bar z_1)\sigma(0,0)\vert 0,0;\alpha_0\rangle} 
{\langle \mathrm{free}\vert 0,0;\alpha_0\rangle}
=0,\\
(\mathrm{new}):&&\frac{\langle \mathrm{new} \vert\sigma(z_1,\bar z_1)\sigma(0,0)\vert 0,0;\alpha_0\rangle} 
{\langle \mathrm{new}\vert 0,0;\alpha_0\rangle}
=0,
\eea
with $I_{\sigma}$ and $I_{\psi}$ given by (\ref{eqn:Isigma}) and (\ref{eqn:Ipsi}). 

The other type of 2-point function $\langle\sigma^\dag \sigma\rangle_{boundary}$ can be computed
 in a similar manner, but this time the $W$ violating boundary $\langle B(\alpha,w_\rho)\vert$ gives non-trivial amplitudes.
There are four conformal blocks:
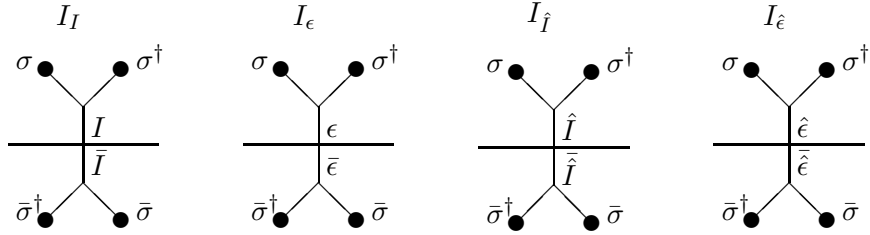
\setlength{\unitlength}{1mm}
\begin{center}
\begin{minipage}{30mm}
\begin{center}
 $I_I$

~~~\begin{picture}(20,30)
{\thicklines
\put(0,15){\line(1,0){20}}}
{\thinlines
\put(10,20){\line(0,-1){10}}
\put(10,20){\line(1,1){5}}
\put(10,20){\line(-1,1){5}}
\put(10,10){\line(1,-1){5}}
\put(10,10){\line(-1,-1){5}}}
\put(5,25){\circle*{2}}
\put(15,25){\circle*{2}}
\put(5,5){\circle*{2}}
\put(15,5){\circle*{2}}
\put(1,25){$\sigma$}
\put(17,25){$\sigma^\dag$}
\put(1,5){$\bar\sigma^\dag$}
\put(17,5){$\bar\sigma$}
\put(11,16){$I$}
\put(11,11){$\bar I$}
\end{picture}
\end{center}
\end{minipage}
\begin{minipage}{30mm}
\begin{center}
 $I_\epsilon$

~~~\begin{picture}(20,30)
{\thicklines
\put(0,15){\line(1,0){20}}}
{\thinlines
\put(10,20){\line(0,-1){10}}
\put(10,20){\line(1,1){5}}
\put(10,20){\line(-1,1){5}}
\put(10,10){\line(1,-1){5}}
\put(10,10){\line(-1,-1){5}}}
\put(5,25){\circle*{2}}
\put(15,25){\circle*{2}}
\put(5,5){\circle*{2}}
\put(15,5){\circle*{2}}
\put(1,25){$\sigma$}
\put(17,25){$\sigma^\dag$}
\put(1,5){$\bar\sigma^\dag$}
\put(17,5){$\bar\sigma$}
\put(11,16){$\epsilon$}
\put(11,11){$\bar\epsilon$}
\end{picture}
\end{center}
\end{minipage}
\begin{minipage}{30mm}
\begin{center}
 $I_{\Ibar }$

~~~\begin{picture}(20,30)
{\thicklines
\put(0,15){\line(1,0){20}}}
{\thinlines
\put(10,20){\line(0,-1){10}}
\put(10,20){\line(1,1){5}}
\put(10,20){\line(-1,1){5}}
\put(10,10){\line(1,-1){5}}
\put(10,10){\line(-1,-1){5}}}
\put(5,25){\circle*{2}}
\put(15,25){\circle*{2}}
\put(5,5){\circle*{2}}
\put(15,5){\circle*{2}}
\put(1,25){$\sigma$}
\put(17,25){$\sigma^\dag$}
\put(1,5){$\bar\sigma^\dag$}
\put(17,5){$\bar\sigma$}
\put(11,16){$\Ibar$}
\put(11,10){$\bar{\Ibar}$}
\end{picture}
\end{center}
\end{minipage}
\begin{minipage}{30mm}
\begin{center}
 $I_{\Ebar}$

~~~\begin{picture}(20,30)
{\thicklines
\put(0,15){\line(1,0){20}}}
{\thinlines
\put(10,20){\line(0,-1){10}}
\put(10,20){\line(1,1){5}}
\put(10,20){\line(-1,1){5}}
\put(10,10){\line(1,-1){5}}
\put(10,10){\line(-1,-1){5}}}
\put(5,25){\circle*{2}}
\put(15,25){\circle*{2}}
\put(5,5){\circle*{2}}
\put(15,5){\circle*{2}}
\put(1,25){$\sigma$}
\put(17,25){$\sigma^\dag$}
\put(1,5){$\bar\sigma^\dag$}
\put(17,5){$\bar\sigma$}
\put(11,16){$\Ebar$}
\put(11,11){$\bar{\Ebar}$}
\end{picture}
\end{center}
\end{minipage}
\end{center}
which are found to be
\bea
I_I
&=&N_{I}\xi^{-2/15}(1-\xi)^{4/15}F(\frac 45, \frac 15, \frac 35;\xi),\\
I_\epsilon
&=&N_{\epsilon}\xi^{4/15}(1-\xi)^{4/15}F(\frac 65, \frac 35, \frac 75;\xi),\\
I_{\Ibar }
&=&N_{\Ibar }\xi^{-2/15}(1-\xi)^{8/15}F(\frac 45, \frac 25, \frac 35;\xi),\\
I_{\Ebar}
&=&N_{\Ebar}\xi^{4/15}(1-\xi)^{8/15}F(\frac 45, \frac 65, \frac 75;\xi).
\eea
The normalization constants are determined as
\bea
&&N_{I}=N_{\Ibar }=1,\\
&&N_{\epsilon}=N_{\Ebar}=C_{\sigma^\dag\sigma}{}^\epsilon
=\frac{\Gamma(1/5)\Gamma(3/5)}{2\Gamma(2/5)^2\lambda},
\eea
where $C_{\sigma^\dag\sigma}{}^\epsilon$ is the 3-point coupling constant in the OPE, 
\bea
&&\sigma^\dag(z_1,\bar z_1)\sigma(z_2,\bar z_2)
=|z_1-z_2|^{-4/15}+C_{\sigma^\dag\sigma}{}^\epsilon |z_1-z_2|^{8/15}\epsilon(w,\bar w)+\cdots.
\eea
The 2-point function $\langle\sigma^\dag\sigma\rangle_{boundary}$ for the eight boundary conditions is,
\bea
(\mathrm{fixed}):&&\frac{\langle\tilde I, \tilde\psi, \tilde\psi^\dag\vert\sigma^\dag (z_1,\bar z_1)\sigma(0,0)\vert 0,0;\alpha_0\rangle}
{\langle\tilde I,\tilde\psi,\tilde\psi^\dag\vert 0,0;\alpha_0\rangle}
=I_{I}+\lambda I_{\epsilon},\\
(\mathrm{mixed}):&&\frac{\langle\tilde\epsilon,\tilde\sigma,\tilde\sigma^\dag\vert\sigma^\dag (z_1,\bar z_1)\sigma(0,0)\vert 0,0;\alpha_0\rangle} 
{\langle\tilde\epsilon,\tilde\sigma,\tilde\sigma^\dag\vert 0,0;\alpha_0\rangle}
=I_{I}-\frac{1}{\lambda^3}I_{\epsilon},\\
(\mathrm{free}):&&\frac{\langle \mathrm{free}\vert\sigma^\dag(z_1,\bar z_1)\sigma(0,0)\vert 0,0;\alpha_0\rangle} 
{\langle \mathrm{free}\vert 0,0;\alpha_0\rangle}
=I_{\Ibar }-\lambda I_{\Ebar},\\
(\mathrm{new}):&&\frac{\langle\mathrm{ new} \vert\sigma^\dag(z_1,\bar z_1)\sigma(0,0)\vert 0,0;\alpha_0\rangle} 
{\langle \mathrm{new}\vert 0,0;\alpha_0\rangle}
=I_{\Ibar }+\frac{1}{\lambda^3}I_{\Ebar}.
\eea

We have seen that the 2-point functions are determined completely, and it is easy to verify that the 
 results have reasonable near-boundary behaviours.
We have checked (by computer) that the six conformal blocks 
$I_{\sigma}$, $I_{\psi}$, $I_{I}$, $I_{\Ibar }$, $I_{\epsilon}$, $I_{\Ebar}$
 all satisfy the 6th order differential equation obtained from (\ref{eqn:singvec6}),
 and also that their Wronskian is non-trivial. 
Thus these conformal blocks are indeed six independent solutions of the differential equation.



\section{Discussion}

The purpose of this paper has been to describe the extension
of the Coulomb gas formulation of boundary CFTs to models with a
higher chiral algebra. We have shown in detail how this works 
for $A_2$ to yield exactly the conformal boundary states that are
expected on the basis of general arguments for the three state Potts
model. At the level of a single boundary (ie a disk topology) the 
correlation functions that we have computed  are all consistent with 
our understanding of the physics of the model. The $\sigma$
 two point amplitudes lead naturally to the two extra conformal blocks
that are expected over and above those associated in the ordinary 
(ie as in rational minimal models) way with the primary fields. 
This is very satisfactory; in this formulation we can calculate an 
integral representation for essentially
any correlation function of the  CFT in the disk topology, just as the 
original Coulomb gas method can for the bulk theory. 

There are some questions that are not settled. Firstly the work reported here 
is not a complete test of the formulation at higher topology. The expectation
 of the identity operator in the cylinder topology is
 given correctly (ie as the cylinder amplitudes) but we have not calculated
other correlation functions and the constants $\kappa_{wMN}$ in 
(\ref{notIshi}) have not been completely determined (at least up to trivial ambiguities).
 This is all intimately related to the BRST formulation of the 
Felder complex for the $W_3$ algebra 
and there seem to be some gaps in what is known about this; for
example, so far as we know, there is no proof of the statement (\ref{coho}) in the
literature. Secondly it is clear that the general structure of this
 formulation can be extended without much difficulty to higher $W$ algebras
and can be used as a means of classifying possible conformal invariant 
boundary states which do not preserve the higher symmetry; whether all such 
states can be found this way is an open question.

\acknowledgments

The support of PPARC grant PPA/G/0/2000/00469
 is acknowledged. We acknowledge useful
conversations with J.B. Zuber and M. Gaberdiel. AFC is supported by FCT
(Portugal) through fellowship SFRH/BD/1125/2000.

\appendix

\section {$W$ conserving screening operators}
 Note that 
$W$ invariance of $Q_{\alpha}$ is equivalent to the statement
\begin{equation} W(z) V_\alpha(w)=\partial_w(\ldots)+{\rm regular~terms}.
\end{equation}
In turn this can be checked by considering whether it is the case that
\begin{equation}\mathcal D_3:e^{i\alpha\cdot\phi(w)}:\,
=\partial_w(\ldots)+{\rm regular~terms}.
\label{Winv}\end{equation}
If so 
then, since we know that the Virasoro generators commute with $Q_\alpha$,
 it would follow that so does $W$.
It is a straightforward but tedious exercise to use Wick's
 theorem to compute the singular parts of the left hand side of (\ref{Winv})
for a general $\alpha$,
\begin{equation}
\alpha=\alpha_1 e_1+\alpha_2 e_2.
\end{equation}
After repeatedly discarding 
total derivatives with respect to $w$ we are left with
\begin{align}
:\frac{(2i\alpha_0)^{-3}}{z-w}\Bigg(&-2\alpha_0\alpha_1 e_1\cdot\partial^2\phi
-\half \alpha_1(4\alpha_0-\alpha_2)(2\alpha_0+\alpha_2-\alpha_1)
\alpha\cdot\partial^2\phi\nonumber\\
&+i\bigg(\alpha_2 h_1\cdot\partial\phi\, e_2\cdot\partial\phi+
\alpha_1 h_3\cdot\partial\phi\,e_1\cdot\partial\phi\nonumber\\
&+\alpha\cdot\partial\phi\bigg(
(2\alpha_0\alpha_2-4\alpha_0\alpha_1-\alpha_2^2+\alpha_2\alpha_1)
h_1\cdot\partial\phi+\alpha_1(\alpha_2-\alpha_1)
h_3\cdot\partial\phi\nonumber\\
&-\alpha_1\alpha_2 h_2\cdot\partial\phi
-\frac{\alpha_1}{2}(4\alpha_0-\alpha_2)
(2\alpha_0+\alpha_2-\alpha_1)
\alpha\cdot\partial\phi\bigg)\bigg)\Bigg): V_\alpha(w).
\end{align}
In order for (\ref{Winv}) to hold this must be zero. The linear terms in
$\phi$ give
\begin{eqnarray}
-2\alpha_0\alpha_1 
-\half \alpha_1^2(4\alpha_0-\alpha_2)(2\alpha_0+\alpha_2-\alpha_1)&=&0,\nonumber\\
-\half \alpha_1 \alpha_2(4\alpha_0-\alpha_2)(2\alpha_0+\alpha_2-\alpha_1)&=&0.
\end{eqnarray}
The first solution is $\alpha_2=0,\,\alpha_1=\alpha_\pm$
where
\begin{equation}
\alpha_\pm^2-2\alpha_0\alpha_\pm=1.
\end{equation}
 The second
solution is $\alpha_1=0$
 and we must then examine the quadratic terms in $\phi$ which show that 
$\alpha_2=\alpha_\pm$. There are no other solutions
so we see that the \emph{only} $W$ conserving
screening operators are 
\begin{eqnarray}
Q^{(1)}_\pm&=&\oint dz:e^{i\alpha_\pm e_1\cdot\phi(z)}:,\nonumber\\
Q^{(2)}_\pm&=&\oint dz:e^{i\alpha_\pm e_2\cdot\phi(z)}:,\end{eqnarray}
and that all the others violate
the $W$ symmetry.

\section{$\Theta$ function identities}

The equivalence of the cylinder amplitudes (\ref{Wamps}) and (\ref{WVir}) is not entirely straightforward to prove. We show
here one way of doing it for the mixed amplitude in the Identity representation case. The techniques
we use are described for example in the  book \cite{Baker}.

Using the standard results for the Virasoro characters in (\ref{WVir}) gives us
\bea R&=&\chi_{1,1}-\chi_{4,1}\nn\\
&=&\frac{q^{-\frac{1}{30}}}{\prod_{k>0}(1-q^k)}\sum_{n\in\allint}
q^{30n^2+n}-q^{30n^2+11n+1}-q^{30n^2+19n+3}+q^{30n^2+29n+7}.
\eea
On the other hand from (\ref{Wamps}) we expect this to be equal to 
\bea L=\frac{q^{-\frac{1}{30}}}{\prod_{k>0}(1-q^{2k})}\sum_{n\in\allint}q^{20n^2+2n}-q^{20n^2+18n+4}.\eea
 First note that
$L$ can be rewritten
\beq L=\frac{q^{-\frac{1}{30}}}{\prod_{k>0}(1-q^{2k})}\sum_{n\in\allint} (-1)^nq^{5n^2-n}\eeq
and then consider
\bea L'&=&q^{\frac{1}{30}}\prod_{k>0}(1-q^k)^2 L\nn\\
&=&\prod_{k>0}(1-q^{2k})(1-q^{2k-1})^2\sum_{n\in\allint} (-1)^nq^{5n^2-n}\nn\\
&=& \sum_{m,n\in\allint} (-1)^m q^{m^2}  (-1)^n q^{5n^2-n},\eea
where in the last step we have used the Jacobi triple product formula.
Similarly we find that 
\bea R'&=&q^{\frac{1}{30}}\prod_{k>0}(1-q^k)^2R\nn\\
&=&\sum_{m,n\in\allint}(-1)^mq^{\threehalves m^2-\half m}
\left(q^{30n^2+n}-q^{30n^2+11n+1}-q^{30n^2+19n+3}+q^{30n^2+29n+7}\right)\nn\\
&=& \sum_{m,n\in\allint}(q^{6m^2-m}-q^{6m^2+5m+1})\nn\\
&&\qquad\qquad\times\left(q^{30n^2+n}-q^{30n^2+11n+1}-q^{30n^2+19n+3}+q^{30n^2+29n+7}\right).\label{AA1}\eea
We wish to prove that $Q\equiv L'-R'=0$.

Note that writing
\beq m=k-l,\quad n=5k+l+p,\label{magix}\eeq
and summing over $k,l\in \allint$ and $p=0,1,2,3,4,5$ is equivalent to summing
over $m,n\in\allint$. Hence
\bea L'&=&\sum_{k,l\in\allint}\,\sum_{p=0}^5 (-1)^p q^{30k^2+6l^2+2p(5k+l)-k+l+p^2},\label{AA2}\eea
and so using (\ref{AA1}) and (\ref{AA2}) we find
\bea Q&=&\sum_{l,k\in \allint}q^{6l^2+l}q^{30k^2+19k+3}
-\sum_{l,k\in \allint}q^{6l^2+3l}q^{30k^2+9k+1}\nn\\
&&+\sum_{l,k\in \allint}q^{6l^2+5l+l}q^{30k^2+k}
-\sum_{l,k\in \allint}q^{6l^2+7l+2}q^{30k^2+11k+1}\nn\\
&&+\sum_{l,k\in \allint}q^{6l^2+9l}q^{30k^2+39k+16}
-\sum_{l,k\in \allint}q^{6l^2+l}q^{30k^2+29k+7}\nn\\
&=&\sum_{k,l\in\allint}\,\sum_{p=0}^5 (-1)^p q^{30k^2+6l^2+(19-10p)k+(2p+1)l+p^2-3p+3}.\label{uhoh}
\eea
Now  change the sign of $k$ in (\ref{uhoh}) and make the change 
of variables (\ref{magix}) in reverse to find that 
\beq Q=\sum_{m,n\in\allint}(-1)^{m+n}q^{5n^2-4n+(m-\threehalves)^2+\threequarters}.\eeq
We see immediately that the power of $q$ in $Q$ is invariant under $m\to 3-m$ but the phase factor
changes sign; hence $Q=0$.

\input{potts090603.bbl}

\end{document}

%% file: potts090603.bbl
\providecommand{\href}[2]{#2}\begingroup\raggedright\endgroup

%% file: potts090603.bbl
\begin{thebibliography}{10}

\bibitem{Dotsenko:1984nm}
V.~S. Dotsenko and V.~A. Fateev, {\it Conformal algebra and multipoint
  correlation functions in 2d statistical models},  {\em Nucl. Phys.} {\bf
  B240} (1984) 312.

\bibitem{Dotsenko:1985ad}
V.~S. Dotsenko and V.~A. Fateev, {\it Four point correlation functions and the
  operator algebra in the two-dimensional conformal invariant theories with the
  central charge $c\leq 1$},  {\em Nucl. Phys.} {\bf B251} (1985) 691.

\bibitem{Cardy:1984bb}
J.~L. Cardy, {\it Conformal invariance and surface critical behavior},  {\em
  Nucl. Phys.} {\bf B240} (1984) 514--532.

\bibitem{Kawai:2002vd}
S.~Kawai, {\it Coulomb-gas approach for boundary conformal field theory},  {\em
  Nucl. Phys.} {\bf B630} (2002) 203--221,
  [\href{http://xxx.lanl.gov/abs/hep-th/0201146}{{\tt hep-th/0201146}}].

\bibitem{Kawai:2002pz}
S.~Kawai, {\it Free-field realisation of boundary states and boundary
  correlation functions of minimal models},  {\em J. Phys.} {\bf A36} (2003)
  6547, [\href{http://xxx.lanl.gov/abs/hep-th/0210032}{{\tt hep-th/0210032}}].

\bibitem{Fateev:1987vh}
V.~A. Fateev and A.~B. Zamolodchikov, {\it Conformal quantum field theory
  models in two-dimensions having $Z_3$ symmetry},  {\em Nucl. Phys.} {\bf B280}
  (1987) 644--660.

\bibitem{Cardy:1989ir}
J.~L. Cardy, {\it Boundary conditions, fusion rules and the Verlinde formula},
  {\em Nucl. Phys.} {\bf B324} (1989) 581.

\bibitem{Affleck:1998nq}
I.~Affleck, M.~Oshikawa, and H.~Saleur, {\it Boundary critical phenomena in the
  three-state Potts model},  {\em J. Phys.} {\bf A31} (1998) 5827,
  [\href{http://xxx.lanl.gov/abs/cond-mat/9804117}{{\tt cond-mat/9804117}}].

\bibitem{Fuchs:1998qn}
J.~Fuchs and C.~Schweigert, {\it Completeness of boundary conditions for the
  critical three-state Potts model},  {\em Phys. Lett.} {\bf B441} (1998)
  141--146, [\href{http://xxx.lanl.gov/abs/hep-th/9806121}{{\tt
  hep-th/9806121}}].

\bibitem{Behrend:1998mu}
R.~E. Behrend, P.~A. Pearce, and J.-B. Zuber, {\it Integrable boundaries,
  conformal boundary conditions and A-D-E fusion rules},  {\em J. Phys.} {\bf
  A31} (1998) L763--L770, [\href{http://xxx.lanl.gov/abs/hep-th/9807142}{{\tt
  hep-th/9807142}}].

\bibitem{Felder:1989zp}
G.~Felder, {\it BRST approach to minimal models},  {\em Nucl. Phys.} {\bf
  B317} (1989) 215.

\bibitem{Bilal:1991eu}
A.~Bilal, {\it Introduction to W algebras}, Presented at the Spring School on
  String Theory and Quantum Gravity, Trieste, Italy, Apr 15-23, 1991.

\bibitem{Bouwknegt:1993wg}
P.~Bouwknegt and K.~Schoutens, {\it W symmetry in conformal field theory},
  {\em Phys. Rept.} {\bf 223} (1993) 183--276,
  [\href{http://xxx.lanl.gov/abs/hep-th/9210010}{{\tt hep-th/9210010}}].

\bibitem{Bouwknegt:1990xa}
P.~Bouwknegt, J.~G. McCarthy, and K.~Pilch, {\it Quantum group structure in the
  Fock space resolutions of $\hat{sl}(n)$ representations},  {\em Commun. Math. Phys.}
  {\bf 131} (1990) 125--156.

\bibitem{Mizoguchi:1992vt}
S.~Mizoguchi and T.~Nakatsu, {\it BRST structure of the $W_3$ minimal model},
  {\em Prog. Theor. Phys.} {\bf 87} (1992) 727--741.

\bibitem{DiFrancesco:1997nk}
P.~Di~Francesco, P.~Mathieu, and D.~Senechal, {\em Conformal Field Theory}.
\newblock Springer, New York, USA., 1997.

\bibitem{Cardy:1991tv}
J.~L. Cardy and D.~C. Lewellen, {\it Bulk and boundary operators in conformal
  field theory},  {\em Phys. Lett.} {\bf B259} (1991) 274--278.

\bibitem{Mizoguchi:1991pf}
S.~Mizoguchi, {\it The structure of representation of the $W_3$ algebra},  {\em
  Int. J. Mod. Phys.} {\bf A6} (1991) 133--162.

\bibitem{Baker}
H.~F. Baker, {\em Abelian Functions}.
\newblock Cambridge University Press, UK., 1897, reissued 1995.

\end{thebibliography}
